\documentclass[usenatbib]{mnras}

\usepackage{mathptmx}
\usepackage[T1]{fontenc}
\usepackage{ae,aecompl}
\usepackage{amsmath}
\usepackage{amssymb}	% Extra maths symbols
\usepackage{graphicx}	% Including figure files
\usepackage{color}
\usepackage[normalem]{ulem}
\usepackage{threeparttable}

\newcommand{\rom}[1]{\MakeUppercase{\romannumeral #1}} %Roman number
\newcommand{\Mcrsb}{M_{\rm oc}^{\rm cr(sb)}}
\newcommand{\comsb}{COM-SB}
\newcommand{\comhp}{COM-HP}

\newcommand{\epsl}{Earth Planet. Sci. Lett.} % Earth and Planetary Science Letters
\newcommand{\cg}{Chem. Geol.} % Chemical Geology
\newcommand{\mc}{Mar. Chem.} % Marine chemistry
\newcommand{\pepi}{Phys. Earth Planet. Inter.} % Physics of the Earth and Planetary Interiors
\newcommand{\nag}{Nature Geosci.} % Nature Geoscience
\newcommand{\jpcm}{J. Phys. Condens. Matter} % Journal of Physics: Condensed Matter
\newcommand{\solsr}{Sol. Syst. Res.} %Solar System Research
\newcommand{\nc}{Nature Commun.} % Nature Communications
\newcommand{\gji}{Geophys. J. Int.} % Geophysical Journal International
\newcommand{\gsf}{Geosci. Frontiers} % Geoscience Frontiers
\newcommand{\ajs}{Am. J. Sci.} % American Journal of Science
\newcommand{\jas}{J. Atmos. Sci.} % Journal of Atmospheric Sciences
\newcommand{\pnas}{Proc. Natl. Acad. Sci. USA} % Proceedings of the National Academy of Sciences of the United States of America

\title[Runaway climate cooling of ocean planets]{
Runaway climate cooling of ocean planets in the habitable zone: a consequence of seafloor weathering enhanced by melting of high-pressure ice}

\author[A. Nakayama et al.]{
A. Nakayama,$^{1}$\thanks{E-mail: anakayama@eps.s.u-tokyo.ac.jp (AN)}
T. Kodama,$^{1,2,3}$
M. Ikoma,$^{1,4}$
and Y. Abe$^{1,5}$
\\
$^{1}$Department of Earth and Planetary Science, Graduate School of Science, The University of Tokyo, 7-3-1 Hongo, Bunkyo-ku, Tokyo 113-0033, Japan \\
$^{2}$Center for Earth surface system dynamics, Atmospheric and Ocean Research Institute, The University of Tokyo, 5-1-5 Kashiwanoha, Kashiwa, Chiba 277-8568, Japan \\
$^{3}$Laboratoire d'{}astrophysique de Bordeaux, Universit\'e de Bordeaux, B18 All\'ee Geoffroy Saint-Hilaire, 33615 Pessac, France \\
$^{4}$Research Center for the Early Universe (RESCEU), Graduate School of Science, The University of Tokyo, 7-3-1 Hongo, Bunkyo-ku, Tokyo 113-0033, Japan \\
$^{5}$Deceased
}

\date{Accepted XXX. Received YYY; in original form ZZZ}
\pubyear{2019}

\begin{document}
\label{firstpage}
\pagerange{\pageref{firstpage}--\pageref{lastpage}}
\maketitle

%=================== Abstract =====================
\begin{abstract}

Terrestrial planets covered globally with thick oceans (termed ocean planets) in the habitable zone were previously inferred to have extremely hot climates in most cases.
This is because ${\rm H_2O}$ high-pressure (HP) ice on the seafloor prevents chemical weathering and, thus, removal of atmospheric CO$_2$.
Previous studies, however, ignored melting of the HP ice and horizontal variation in heat flux from oceanic crusts.
Here we examine whether high heat fluxes near the mid-ocean ridge melt the HP ice and thereby remove atmospheric ${\rm CO_2}$.
We develop integrated climate models of an Earth-size ocean planet with plate tectonics for different ocean masses, which include the effects of HP ice melting, seafloor weathering, and the carbonate-silicate geochemical carbon cycle.
We find that the heat flux near the mid-ocean ridge is high enough to melt the ice, enabling seafloor weathering.
In contrast to the previous theoretical prediction, we show that climates of terrestrial planets with massive oceans lapse into extremely cold ones (or snowball states) with CO$_2$-poor atmospheres.
Such extremely cold climates are achieved mainly because the HP ice melting fixes seafloor temperature at the melting temperature, thereby keeping a high weathering flux regardless of surface temperature.
We estimate that ocean planets with oceans several tens of the Earth's ocean mass no longer maintain temperate climates.
These results suggest that terrestrial planets with extremely cold climates exist even in the habitable zone beyond the solar system, given the frequency of water-rich planets predicted by planet formation theories.

\end{abstract}


% %*************************************** Introduction **********************************************

	%*************************************** Introduction **********************************************

\section{Introduction}
The Earth's climate system is generally thought to be stabilized by a carbonate-silicate geochemical cycle of carbon (hereafter called the \textit{carbon cycle}).
On geological timescales, the amount of the greenhouse gas ${\rm CO_2}$ is determined by a balance between degassing flux through volcanism and sinking flux through chemical weathering.
Since chemical weathering becomes more efficient with temperature, a negative-feedback mechanism operates to keep the CO$_2$ partial pressure at low levels and, thus, to maintain the temperate climate \citep{Wal1981}.
In the present Earth, weathering occurs mainly on continents \citep[]{Cal1995}.

Beyond the solar system, however, there must be continent-free terrestrial planets completely covered with oceans in the habitable zone.
%\naka{The habitable zone is defined as the circumstellar region that the ocean exists on the planetary surface.}
In this study we refer to the conventional habitable zone defined based on 1-D radiative-convective models, ranging from 0.34 to 1.06 times the present solar insolation at the Earth's orbit \citep[]{Kas1993, Kop2013}, as the habitable zone.
Given diverse water-supply processes and their stochastic nature, terrestrial exoplanets must be diverse in ocean mass.
Indeed, many recent theories of planet formation predict that terrestrial exoplanets could have much more water than the Earth (see recent reviews by \citet{OBrien2018} and \citet{Iko2018}).
$N$-body simulations of late-stage terrestrial planet accretion including the supply of water-rich planetesimals beyond the snowline demonstrate that terrestrial planets with oceans of ten to several hundred Earth's ocean masses might be common in the habitable zone \citep[]{Ray2004, Ray2007}. 
On the Earth, there would be no lands if the ocean mass were three times larger than the present (i.e., 0.023~\% of the Earth's mass) \citep[]{Mar2013,Kod2018}.

What is the climate like on terrestrial planets completely covered with oceans and what
influence does ocean mass have on the climate?
Such terrestrial planets are called \textit{ocean planets}, hereafter, whereas
ones covered partially with oceans like the Earth are called \textit{partial ocean planets} \citep[]{Kun2003, Leg2004}.
On ocean planets, seafloor weathering, instead of continental weathering, would control the planetary climate.
The role of seafloor weathering in climate is of interest in this study.
In particular, we focus on the influence of high-pressure (HP) ice of ${\rm H_2O}$ such as ice~\rom{6} and \rom{7} on the seafloor weathering.

Planets with larger water amounts than a certain threshold have the HP ice on the seafloor, provided the ocean has a steady, isothermal or adiabatic structure \citep[see Fig.~2 of][]{Leg2004}.
Since the HP ice is a solid heavier than its counterpart liquid, a solid layer is formed between the ocean and oceanic crust
and prevents seafloor weathering \citep[]{Ali2014, Kal2013}.
% 	Then, the amount of atmospheric CO$_2$ is regulated by the solubility of ${\rm CO_2}$ in the ocean.
% 	\citet[]{Wor2013b} found a positive feedback mechanism such that a rise in surface temperature reduces the solubility of ${\rm CO_2}$ in the ocean and, thus, increases the atmospheric CO$_2$, enhancing the greenhouse effect.

Climates of ocean planets without geochemical interaction and carbon cycle between the ocean-atmosphere system and silicate mantle were previously investigated. 
\citet[]{Wor2013b} and \citet[]{Kit2015} explored the effect of dissolution of ${\rm CO_2}$ into a cation-poor ocean and found that the CO$_2$ pressure decreases with increasing temperature for a given carbon inventory in the atmosphere-ocean system (see Fig.~2 of \citet{Kit2015}).
This suggests that such a climate system is an unstable one with a positive feedback cycle. 
\citet[]{Kit2018} considered supply of cations to the ocean, which strongly affects ocean chemistry, in the initial, hot stage after solidification of the magma ocean.
They showed that large cation concentration enhances the positive feedback and leads to destabilizing planetary climate into hot one for a large ${\rm CO_2}$ inventory ($\sim$ 100~bars) in the atmosphere-ocean system even for stellar insolation comparable to the present Earth.

However, whether the layer of HP ice really exists and prevents seafloor weathering completely must be verified through a detailed consideration of heat transfer and rheology in the HP ice layer.
\citet[]{Noa2016} examined the stability of the HP ice layer by performing non-steady, one-dimensional simulations of heat transfer, including the melting of HP ice, in the layer (liquid H$_2$O + HP ice) above the oceanic crust (collectively called the \textit{H$_2$O layer,} hereafter).
They found that the heat flux from the oceanic crust is too high for steady heat transport in the HP ice and, thus, the heat is temporarily stored near the bottom of the ${\rm H_2O}$ layer, which results in melting the HP ice.
Since the resultant melt is lighter than its surroundings, an upwelling flow of partially molten HP ice occurs.
Such a possibility has been investigated also in studies of large icy moons in the solar system, in particular, Ganymede, which propose that solid and liquid coexist via melt production within the HP ice layer, bringing about a melt-buoyancy-driven upwelling flow in the interior.

To evaluate the efficiency of heat transport by the melt-buoyancy-driven flow,
\citet[]{Cho2017} performed 3-D simulations of thermal convection in the HP ice layer, including the effect of melting of the HP ice.
In their simulations,
they assumed and mimicked a permeable flow in the HP ice by extracting the generated melt instantaneously to the above ocean.
Then, they demonstrated that melt is mostly generated on the oceanic crust and the permeable flow dominates the heat transport.
Recently, \citet[]{Kal2018} performed 2-D convection simulation of a water-ice mixture to investigate the behavior of the generated melt in the HP ice layer.
They demonstrated that heat is efficiently transported by the melt-buoyancy-driven
convective and permeable flows and water is exchanged throughout the HP ice layer.
In this study, we call those flows the \textit{sorbet flow}, since they are flows of a water-ice mixture.
The sorbet flow occurs for the small thickness of the HP ice ($\lesssim 200$~km) and large heat flow ($\gtrsim 20$~mW~m$^{-2}$) for Ganymede-like icy bodies.
Nusselt--Rayleigh number scaling supports that such a sorbet flow likely occurs also for ocean planets with Earth-like geothermal heats ($80$~mW~m$^{-2}$ in the present Earth's mean mantle heat flow) and thicker HP ice.
Hence, seafloor weathering likely occurs for ocean planets with the HP ice.

Horizontal variation is another important effect ignored previously.
In particular, for planets where plate tectonics works, the heat flow from oceanic crusts is highest at mid-ocean ridges and decreases with distance from there.
The heat flow near mid-ocean ridges can be high enough to melt the HP ice.
Then, the seafloor temperature is fixed close to the melting temperature for the pressure at the seafloor (hereafter, the seafloor pressure).
This temperature is much higher than one obtained from inward integration of the adiabat from the oceanic surface to the seafloor.
Higher seafloor temperature results in more efficient seafloor weathering, according to the temperature dependence of seafloor weathering inferred based on dissolution experiments of basalt \citep[]{Bra1997, Gud2011} and geological evidence \citep[]{Coo2015, Kri2017}.
Hence, the seafloor weathering can remove atmospheric ${\rm CO_2}$ efficiently, provided such a molten region is sufficiently wide.

This study is aimed at evaluating the role of the HP ice
in seafloor weathering and climate for ocean planets with
a focus on the effects of the liquid-solid coexistence region maintained by the sorbet flow and the horizontal variation in heat flux from the oceanic crust.
The rest of this paper is organized as follows:
In section 2, we describe our model to simulate the ocean layer structure and planetary climate.
In section 3, we show the behavior of the HP ice with a focus on the area where melting occurs.
In section 4, we show the impacts of seafloor weathering with the HP ice on the planetary climate.
In section 5, we discuss surface environments of ocean planets, caveats of the model, and implication of our results for terrestrial exoplanets.
In section 6, we conclude this study.

%*************************************** Introduction **********************************************

% % %*************************************** Introduction **********************************************

% % %*************************************** Methods **********************************************

%*************************************** Methods **********************************************
\section{Climate model}
We consider an Earth-size ocean planet with various amounts of ${\rm H_2O}$ and ${\rm CO_2}$.
Of special interest in this study is the impact of ocean mass, $M_\mathrm{oc}$, on the planetary climate including surface temperature, $T_\mathrm{s}$, and CO$_2$ partial pressure, $P_\mathrm{CO_2}$.
We assume that the planet is almost Earth-like, namely, a terrestrial planet with the Earth's mass and internal composition orbiting at 1~AU far from a Sun-like star, except for the ocean mass.
Our climate model consists of four components:
(1) internal structure integration that determines the thickness of the HP ice layer (section~\ref{sec:OSM});
(2) seafloor environment modeling that determines the area where seafloor weathering works when the HP ice is present (section~\ref{sec:SEM});
(3) carbon cycle modeling that calculates $P_\mathrm{CO_2}$ (section~\ref{sec:carbon});
(4) atmospheric modeling that calculates $T_\mathrm{s}$ (section~\ref{sec:AM}).

%===============================================
%% Ocean structure model
\subsection{Ocean structure model} \label{sec:OSM}

The hydrostatic structure of the ocean is determined by
\begin{eqnarray}
	\frac{dP}{dr} &=& - g \rho, \label{hy}\\
	\frac{dm}{dr} &=& 4 \pi r^2 \rho, \label{mass}
\end{eqnarray}
where $r$ is the radial distance from the planetary center,
$P$ and $\rho$ are the pressure and density, respectively,
$m$ is the cumulative mass, and $g$ is the gravity ($g$ = $G m / r^2$; $G$ being the gravitational constant).
Its thermal structure is assumed to be adiabatic:
	\begin{equation}
	 \frac{dT}{dr} = - \frac{\alpha g T}{C_P}, \label{adiabatic}
	\end{equation}
where $T$ is the temperature, $\alpha$ is the thermal expansivity, and $C_P$ is the heat capacity.

Two of the three boundary conditions are
$T = T_{\rm s}$ and $P = P_{\rm s}$ at $r = R_{\rm p}$, where $P_{\rm s}$ is the surface pressure and $R_{\rm p}$ is the planet radius.
Here $P_{\rm s}$ is the sum of the background pressure ($P_{\rm n}$ = $1 \times 10^5$~Pa) and vapor pressure $P_{\rm H_2O}$ for $T_{\rm s}$, which is taken from \citet[]{Nak1992}; $P_{\rm CO_2}$ is relatively small.
The inner boundary condition is $m$ = 0 at $r$ = 0.
This means that we must continue the integration until the planet's center,
although we are interested only in the ocean layer.
The planet consists of a H$_2$O ocean layer, a rocky mantle, and an iron core.
Regarding the equations of state for the materials and chemical phases, we mostly follow \citet[]{Val2007}.
The details are given in Appendix~\ref{sec:ISM}.

The red line of Fig.~\ref{Figa1} shows the calculated relationship between the surface temperature and critical ocean mass beyond which HP ice exists (see section~\ref{sec:putting} for the numerical procedure), which is abbreviated to \comhp\, hereafter.
It turns out that HP ice exists for an Earth-like planet with an ocean of more than $\sim$20 to $\sim$100~$M_\mathrm{oc, \oplus}$, depending on surface temperature.
To see the sensitivity to the thermal structure of the ocean, we also show the result for an isothermal ocean.
The difference in \comhp\, between the isothermal and adiabatic cases is $\sim1$--$30 M_{\rm oc, \oplus}$ for $T_{\rm s} = $ 280--400~K.
Even for the two extreme cases, the difference is small enough not to change our conclusions.

\begin{figure}
	\includegraphics[width=\columnwidth]{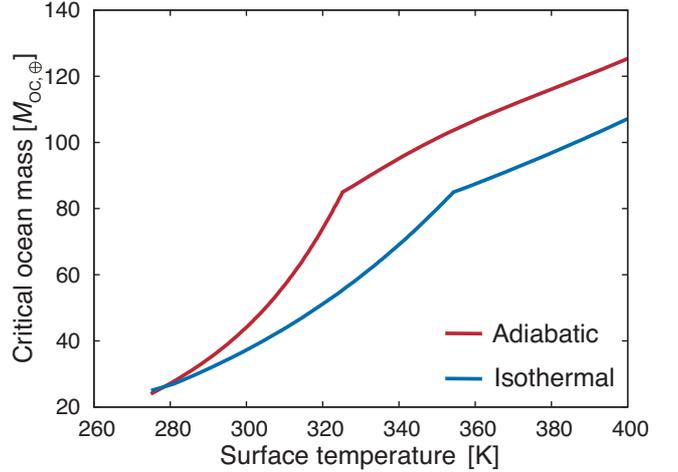}
	\caption{
	The critical ocean mass (\comhp) in the unit of the Earth's ocean mass ($M_{\rm oc, \oplus}$), beyond which high-pressure (HP) ice appears deep in the ocean is shown as a function of surface temperature (red line).
	Note that the result for an isothermal liquid ocean is also shown (blue line) to confirm that our calculation reproduces the result of \citet[]{Kit2015} well.
	\label{Figa1}}
\end{figure}

%===============================================
%% Seafloor environment model
\subsection{Seafloor environment model} \label{sec:SEM}
Near a mid-ocean ridge, heat flow from below is so high that the HP ice would be incapable of transporting the heat by thermal conduction nor convection and consequently become molten.
If liquid water exists together with ice, the heat can be transported efficiently by a sorbet flow, as described in Introduction.
The HP ice far from a mid-ocean ridge remains solid because of low heat flow.
Hence, there is a \textit{critical distance} beyond which or a \textit{critical heat flow,} $q_{\rm cr}$, below which the HP ice remains solid.

\begin{figure*}
\includegraphics[width=17cm]{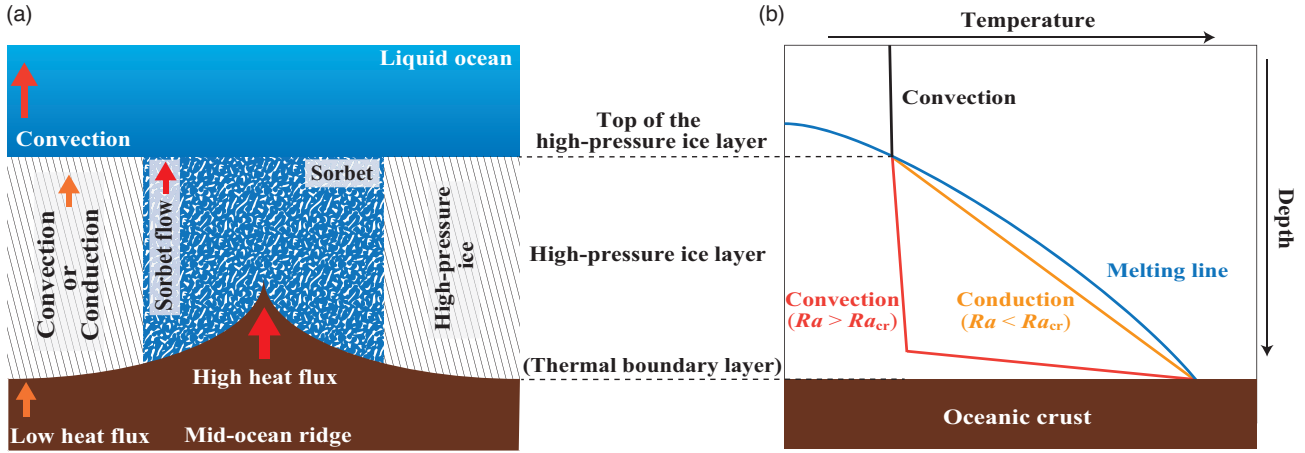}
\caption{
  Seafloor environment model---($a$) schematic illustration of the seafloor environment and
  ($b$) qualitative temperature profiles in the infinitesimally thin layer on the side of the high-pressure ice at the boundary between the "sorbet" and "high-pressure ice" regions.
  In panel~($a$), arrows represent the direction and dominant mechanisms of heat transport.
  In the sorbet region, ice and liquid coexist and thus the temperature is fixed at the melting point of ${\rm H_2O}$.
  In panel~($b$), $Ra$ and $Ra_{\rm cr}$ represent the Rayleigh number and the critical Rayleigh number, respectively.
  The red and orange solid lines represent thermal structure when the high-pressure ice layer is convective and conductive, respectively.
  The blue and black solid lines represent the melting line of ${\rm H_2O}$ and adiabatic thermal structure of the liquid ocean, respectively.
  \label{fig1}}
\end{figure*}

A schematic illustration of our seafloor environment model is shown in Fig.~\ref{fig1}.
Here we assume that
(1) the heat transport is steady and vertically one dimensional,
(2) the composition and phase of H$_2$O are vertically homogeneous in the HP ice region, and
(3) the sorbet flow dominates the heat transport in the solid-liquid coexistence region (called the \textit{sorbet region}, hereafter), whereas solid-state convection or conduction occurs in the HP ice region.
These assumptions are consistent with results of previous hydrodynamical simulations \citep[]{Cho2017, Kal2018}.
We discuss their validity and impact on our conclusion in section~\ref{sec:uHP}.

\subsubsection{Critical heat flow}
First, we determine the critical distance or critical heat flow.
Namely, according to its definition, we find the point at which convection nor conduction can hardly transport the heat inside the HP ice region.
Figure~\ref{fig1}$b$ shows qualitative temperature profiles in the HP ice:
At the critical distance, since the ice-liquid mixture on the oceanic crust is in phase equilibrium,
the temperature is equal to the melting temperature.
Also, the temperature at the top of the HP layer is the melting one, by definition.

To determine the thermal structure of the HP ice region, we adopt a similar approach with that used by \citet[]{Fu2009} who investigated the structure of the icy mantle of an ocean planet with a frozen surface, although they ignored horizontal variation in heat flux.
Unlike \citet{Fu2009}, we take into account the case where the HP ice layer is wholly conductive, ignore the upper thermal boundary layer, and consider the different boundary condition for the bottom of the HP layer. The details are described below.

The mechanism of heat transport depends on the Rayleigh number, $Ra$, which is defined as \citep[]{Tur2002}
\begin{equation}
	 Ra = \frac{g\alpha \rho D^3 \Delta T_{\rm HP}}{\kappa \eta} \label{Ray},
\end{equation}
where $D$ is the thickness of the HP ice layer,
$\kappa$ is the coefficient of thermal diffusivity, $\eta$ is the viscosity,
$\Delta T_{\rm HP} = T^{\rm mel}_{\rm BB} - T^{\rm mel}_{\rm TB}$ and $T^{\rm mel}_{\rm TB}$ and $T^{\rm mel}_{\rm BB}$ are the melting point temperatures for the pressures at the top and bottom of the HP ice layer, respectively.
For the melting point temperature $T^{\rm mel}$, we use the formula from \citet{Dun2010},
\begin{equation} \label{eq:Tmel}
		T^{\rm mel} = a_1 + a_2 P + a_3 \ln{P} + a_4 P^{-1} + a_5 \sqrt{P},
\end{equation}
where $P$ is the pressure in bar and the values of coefficients are summarized in Table~\ref{Table:mel}.
We assume that the phase transition from ice~\rom{6} to \rom{7} occurs at the triple point of liquid/ice~\rom{6}/ice~\rom{7}, the pressure of which is 22160 bars.
The thermal diffusivity is defined by $\kappa = k / \rho C_P$, where $k$ is the thermal conductivity.
For $C_P$ of ice~\rom{6} and \rom{7}, we use the expression derived by \citet[]{Fei1993}.
For $k$, we adopt a constant value of 3.8 ${\rm W m^{-1} K^{-1}}$, which is its typical value for ice~\rom{7} under 2.5~GPa and 300~K \citep[]{Che2011}, for simplicity.
For $\eta$ of ice~\rom{7}, which is poorly constrained,
we adopt a dislocation model for the viscosity of phase~\rom{6}, which is the highest phase of the HP ice measured so far \citep[]{Dur1997} :
\begin{equation}
	 \eta (P_\eta, T_\eta) = B \zeta^{-3.5} \exp \left[ \frac{(E^* + P_\eta V^*)}{R T_\eta}\right],  \label{viscosity}
\end{equation}
where $B$ ($= 6.7 \times 10^{19}$~Pa$^{4.5}$~s) is a constant, $\zeta$ ($= 2.0 \times 10^6$~Pa) is a characteristic shear stress \citep[]{Fu2009}, $R$ is the ideal gas constant, $E^*$ ($= 110$~kJ~mol$^{-1}$) and $V^*$ ($= 1.1 \times 10^{-5}$~m$^3$~mol$^{-1}$) are the activation energy and volume \citep[]{Dur1997}, respectively, and $T_\eta$ and
$P_\eta$ are the temperature and pressure at deformation, respectively.
Because the viscosity contrast in the HP ice layer is relatively small, the small viscosity contrast prescription can be used \citep[]{Fu2009}.
For $P_\eta$ and, $T_\eta$, we use the averaged values for the HP ice layer \citep[]{Dum1999}.
In this study, we assume the value of the critical Rayleigh number, $Ra_\mathrm{cr}$, is 2000.

	\begin{table}
		\begin{center}
			\caption{Coefficients for ice melting curve given by Eq.~(\ref{eq:Tmel}) from \citet[]{Dun2010} \label{Table:mel}}
			\begin{tabular}{cccccc} \hline
				Ice phase      & $a_1$    & $a_2$   & $a_3$    & $a_4$   & $a_5$\\ \hline
				Ice \rom{6} & 4.2804   & -0.0013 & 21.8756  & 1.0018  & 1.0785 \\
				Ice \rom{7} & -1355.42 & 0.0018  & 167.0609 & -0.6633 & 0 \\
				\hline
			\end{tabular}
		\end{center}
	\end{table}

When $Ra < Ra_{\rm cr}$, conduction dominates heat transport and, thus, $q_{\rm cr}$ is given as
	\begin{equation}
	 q_{\rm cr} = k \frac{\Delta T_{\rm HP}}{D}. \label{conduction}
	\end{equation}

When $Ra > Ra_{\rm cr}$, since convection occurs, we assume the adiabatic temperature gradient (i.e., Eq.[\ref{adiabatic}]).
Near physical boundaries, however, since convective motion is prevented, thermal boundary layers are formed, where conduction transports heat.
In this study, we consider the presence of a boundary layer only on the bottom of the HP ice layer (BBL), where the temperature gradient is given by
	\begin{equation}
	\frac{dT}{dr} = - \frac{q}{k}
	\label{conductive}
	\end{equation}
and $q$ is the heat flux.
Integrating Eq.~(\ref{adiabatic}) inwards from the top of the HP ice layer and
Eq.~(\ref{conductive}) outwards from the surface of the oceanic crust,
we determine the BBL's thickness, $\delta$, and the temperature difference in the BBL, $\Delta T_\mathrm{BBL}$ at the crossover point for a given $q$ (see Fig.~\ref{fig1}$b$).

Given that the BBL is marginally stable against convection, $Ra = Ra_\mathrm{cr}$ in the BBL, namely, $Ra_\mathrm{cr} = g \alpha \rho \delta^3 \Delta T_\mathrm{BBL} / \kappa \eta_\mathrm{BBL}$, which comes to be
  \begin{equation}
	\delta = \left( \frac{\kappa \eta_{\rm BBL} Ra_{\rm cr}}{g \alpha \rho \Delta T_{\rm BBL}} \right)^{1/3},
	   \label{TBBL}
  \end{equation}
where $\eta_{\rm BBL}$ is the viscosity of the HP ice in the BBL and calculated with the intermediate values of temperature and pressure between the top and bottom of the BBL.
If the set of $\delta$ and $\Delta T_{\rm BBL}$ for a given $q$ satisfies Eq.~(\ref{TBBL}), the value of $q$ corresponds to $q_{\rm cr}$, which is also written as
	\begin{equation}
	 q_{\rm cr} = k \frac{\Delta T_{\rm BBL}}{\delta} =
	 \frac{k \kappa Ra_{\rm cr}}{g \alpha \rho} \cdot
	 \frac{\eta_{\rm BBL}}{\delta^4}.
	 \label{q_cr}
 	\end{equation}
Note that \citet[]{Fu2009} considered a boundary layer under the top of the HP ice layer in addition to BBL.
We discuss the difference in temperature structure in the HP ice layer between this study and \citet[]{Fu2009} and its impacts on our conclusion in section~\ref{sec:uHP}.

\subsubsection{Effective weathering area}
Same as in the Earth,
the oceanic crust is assumed to form via eruption of hot mantle rock only at the mid-ocean ridge.
As it moves away from the mid-ocean ridge toward the trench, the oceanic crust is cooled by seawater.
Here we define a non-dimensional \textit{effective weathering area}, $f_{\rm oc}$, as the area of the sorbet region (i.e., $q > q_\mathrm{cr}$) relative to the whole area of oceanic crust.
A constant rate of oceanic crust production being assumed,
$f_{\rm oc}$ is equivalent to the ratio of the period during which $q$ $\geq$ $q_{\rm cr}$ to the residence time of the oceanic crust, $\tau$.

To calculate $f_\mathrm{oc}$, we model the cooling of the oceanic crust, adopting the semi-infinite half-space cooling model \citep[]{Tur2002}:
This model assumes that the crust cools only by vertical heat conduction.
The heat flux from the oceanic crust is given by \citep[]{Tur2002}
	\begin{equation}
	 q(t) = \frac{k_{\rm rock} (T_{\rm sol} - T_{\rm floor})}{\sqrt{\pi \kappa_{\rm rock} t}}
	 \equiv \frac{\mathcal{A}}{\sqrt{t}}, \label{semiinf}
	\end{equation}
where $t$ is time, $k_{\rm rock}$ ($= 3.3$~W~m$^{-1}$~K$^{-1}$) and $\kappa_{\rm rock}$ ($= 1.0 \times 10^{-6}$~m$^2$~s$^{-1}$) are the thermal conductivity and thermal diffusivity of the oceanic crust, respectively, $T_{\rm floor}$ is the seafloor temperature, and $T_{\rm sol}$ is the potential temperature of the mantle, for which we assume the peridotite dry solidus at the seafloor pressure, which was parameterized by \citet[]{Hir2009}.
This assumption is made just for simplicity.
The influence of the assumption on planetary climate is discussed in section~\ref{sec:DPC}.

From Eq.~(\ref{semiinf}), the length of time required for $q$ to decrease to $q_\mathrm{cr}$, which is denoted by $t_\mathrm{cr}$, is given by $t_\mathrm{cr} = \mathcal{A}^2/q_\mathrm{cr}^2$.
Also, if the mean mantle heat flow, $\bar{q}$,  is defined as $\bar{q} \equiv \tau^{-1} \int_0^\tau q \, \mathrm{d}t$,
the residence time $\tau$ is given as a function of $\bar{q}$ as $\tau = 4 \mathcal{A}^2 / \bar{q}^2$.
Thus, the effective weathering area is given as
	\begin{equation}
	 f_{\rm oc} \equiv \frac{t_{\rm cr}}{\tau} = \frac{1}{4} \left( \frac{\bar{q}}{q_{\rm cr}}\right)^2 . \label{criticalheat}
	\end{equation}
In some cases, calculated $t_\mathrm{cr}$ happens to be larger than $\tau$, which means the oceanic crust is fully covered with the solid-liquid mixture (i.e., the sorbet).
In such cases, we set $f_{\rm oc} = 1$.
From Eq.~(\ref{criticalheat}), it turns out that when $q_{\rm cr} > {\bar q}/2$, solid HP ice appears near the trench.
In the next section, we use $f_{\rm oc}$ in the carbon cycle model.

In this study, the mean mantle heat flow $\bar{q}$ is a free parameter.
As the fiducial value, we use $\bar{q} = 80$~mW~m$^{-2}$, which is the value for the present Earth.
Note that $q_\mathrm{cr}$ is independent of $\bar{q}$, according to Eq.~(\ref{criticalheat}).

  %===============================================
  %% Carbon cycle model
\subsection{Carbon cycle model} \label{sec:carbon}
In order to investigate planetary climate, we develop a carbon cycle model by modifying the Earth's carbon cycle model of \citet[]{Taj1992}.
Since we focus on continent-free terrestrial planets, we add the effect of seafloor weathering and neglect the continental reservoir of carbon and the effect of continental weathering.
In addition, we consider the presence of the HP ice and pressure-dependent degassing.
Same as \citet{Taj1992} and \citet{Sle2001}, we perform box-model calculations of carbon circulation among reservoirs and find the equilibrium states.

\subsubsection{Carbon reservoirs}
We consider four reservoirs, which include the atmosphere, ocean (liquid water plus HP ice), oceanic crust (basalt), and mantle.
Between the atmosphere and ocean, however, the carbon partition is assumed to be always in equilibrium, which is described in detail in Appendix~\ref{sec:patitioning}.
The equilibrium value of the CO$_2$ partial pressure $P_\mathrm{CO_2}$ depends on the number of cations dissolved in the ocean \citep[e.g.,][]{Zee2001}, for which we assume the present Earth's value, although the supply of cations via continental weathering never occurs in ocean planets.
We have confirmed that overall results are insensitive to the number of dissolved cations (even in the case with no cations in the ocean). This is because the ocean reservoir is much small relative to the whole planetary carbon reservoir.

Carbon dissolved in the ocean is carried to the seafloor in the form of CO$_2$ ice, into which aqueous CO$_2$ is converted in the sorbet region \citep[][]{Bol2013}.
We assume that the carbon circulation in the sorbet region occurs quickly enough that it never affects the mass balance and also planetary climate.
Detailed discussion of the ${\rm CO_2}$ circulation is given in section~\ref{sec:uHP}.

The origin of volatiles in terrestrial planets has been highly debated so far, even for the Earth \citep[e.g.,][]{OBrien2018}.
Since possible candidates such as carbonaceous chondrites and comets include both carbon compounds and water,
we assume that the total mole number of carbon contained in the whole planet, $C_{\rm total}$, is proportional to ocean mass, namely
  \begin{equation}
    C_{\rm total} = \gamma \, n_{\rm oc, \oplus} \frac{M_{\rm oc}}{M_{\rm oc, \oplus}}, \label{Ctotal}
  \end{equation}
where $\gamma$ is the ${\rm CO_2 / H_2O}$ molar ratio in the source of volatiles of the planet, $n_{\rm oc, \oplus} (= 7.6 \times 10^{22}$~mol) is the molar quantity of ${\rm H_2O}$ in the Earth ocean mass, and $M_{\rm oc, \oplus}$ (= $1.37 \times 10^{21}$~kg) is the Earth's ocean mass.
Using the data and estimation published, we can estimate $\gamma $ is to be 0.22 for carbonaceous chondrites \citep[]{Jar1990},  0.71 for comae of comets \citep[]{Mar2016}, and 0.19 for Earth composition \citep[]{Taj1992}.
We use the Earth-like value ($\gamma = 0.19$) as the fiducial value.
Dependence of planetary climate on $\gamma$ is discussed in sections~\ref{sec:ESW}.

\subsubsection{Carbon budget}
The mass balance among those reservoirs is expressed as
  \begin{eqnarray}
    \frac{d (C_{\rm atm} + C_{\rm oc})}{dt} &=& F_{\rm D} + F_{\rm M} - F_{\rm SW}, \label{c_atoc} \\
    \frac{d C_{\rm bs}}{dt} &=& F_{\rm SW} - F_{\rm R} - F_{\rm M}, \label{c_bs} \\
    \frac{d C_{\rm man}}{dt} &=& F_{\rm R} - F_{\rm D},  \label{c_man} \\
    C_{\rm total} &=& C_{\rm atm} + C_{\rm oc} + C_{\rm bs} + C_{\rm man}, \label{c_total}
  \end{eqnarray}
where $C_\mathrm{atm}$, $C_\mathrm{oc}$, $C_\mathrm{bs}$, and $C_\mathrm{man}$ are the mole numbers of carbon contained in the atmosphere, ocean, oceanic basalt, and mantle, respectively, and $F_{\rm SW}, F_{\rm D}, F_{\rm R}, F_{\rm M}$ are the carbon fluxes due to seafloor weathering, degassing from the mid-ocean ridge, regassing via subduction into mantle and metamorphism that leads to degassing from volcanic arc, respectively.
Those equations are solved for a given value of $C_\mathrm{total}$.

We adopt the degassing, regassing, and metamorphism models from \citet[]{Taj1992}, where each flux is expressed as
  \begin{eqnarray}
    F_{\rm D} &=& K_{\rm D} A_{\rm S} C_{\rm man}, \label{degassing}	\\
    F_{\rm R} &=& \frac{\beta}{\tau} C_{\rm bs}, \\
    F_{\rm M} &=& \frac{1 - \beta}{\tau} C_{\rm bs}.   \label{metamorphism}
  \end{eqnarray}
Here $K_{\rm D}$ is the molar fraction of carbon degassing as ${\rm CO_2}$ from the erupting magma per unit area.
We take into account the dependence of $K_\mathrm{D}$ on seafloor pressure (i.e., ocean mass), the detail of which is described in Appendix~\ref{sec:DRM}.
$\beta$ is the regassing ratio defined as the molar fraction of carbonate regassed into the mantle in the total subducting carbonate.
We adopt the present Earth's value of $\beta$ (= 0.4) estimated by \citet[]{Taj1992}.
$A_\mathrm{S}$ is the seafloor spreading rate, which is simply given by
  \begin{equation}
    A_{\rm S} = \frac{A_0}{\tau}, \label{SR}
  \end{equation}
where $A_0$ is the whole area of the seafloor.
We assume that $A_0$ is the present Earth's value ($=3.1 \times 10^{14}$~m$^2$) from \citet[]{McG1989} and calculate $\tau$ from the relation $\tau$ = $4 \mathcal{A}^2 / \bar{q}^2$ for a given $\bar{q}$.

The seafloor weathering rate $F_{\rm SW}$ depends on seafloor temperature $T_{\rm floor}$ as \citep[]{Bra1997}
  \begin{equation}
    F_{\rm SW} = F^*_{\rm SW} f_{\rm oc} \exp \left[ \frac{E_{\rm a}}{R} \left(\frac{1}{T_0} - \frac{1}{T_{\rm floor}} \right) \right] , \label{SW}
  \end{equation}
where $F^*_{\rm SW}$ is the present Earth's seafloor weathering rate,
$f_\mathrm{oc}$ is the effective weathering area given by Eq.~(\ref{criticalheat}),
$E_{\rm a}$ is the activation energy, and $T_0$ (= 289~K) is the reference seafloor temperature that corresponds to the surface temperature obtained by the atmospheric model with the present Earth's condition.
$F^*_{\rm SW}$ estimated from deep-sea cores is $1.5$--$2.9 \times 10^{12}$~mol~yr$^{-1}$ \citep[]{Alt1999, Sta1989, Gil2011}.
In this study, we use $F^*_{\rm SW}$ = $ 2.0 \times 10^{12}$~mol~yr$^{-1}$.

The activation energy $E_{\rm a}$ is uncertain and its reported value ranges between 30 and 92~kJ~mol$^{-1}$.
\citet[]{Bra1997} firstly determined $E_{\rm a}$ experimentally to be 41~kJ~mol$^{-1}$.
Recent inversion methods using geological evidence support a relatively high value of $E_{\rm a}$:
Precisely, strontium and oxygen isotopes in carbonates indicated $E_{\rm a} = 92 \pm 7$~kJ~mol$^{-1}$ \citep[]{Coo2015}.
Also, several proxies reflecting the surface and seafloor temperatures, atmospheric ${\rm CO_2}$, and oceanic pH showed $E_{\rm a} = 75^{+22}_{-21}$~kJ~mol$^{-1}$ \citep[]{Kri2017}.
Those values are also consistent with estimates from laboratory experiments for the dominant minerals in the oceanic crust \citep[]{Bra2014}.
In contrast, an experimental study of basalt dissolution in the moderate pH range reported the relatively small $E_{\rm a}$ of 30~kJ~mol$^{-1}$ \citep[]{Gud2011}.
In this study, we use $E_{\rm a} = 41$~kJ~mol$^{-1}$ as the fiducial value according to previous studies \citep[e.g.,][]{Fol2015} and vary it over the range between 30 and 92~kJ~mol$^{-1}$.
We ignore the pH dependence of seafloor weathering since it is known to be small in the pH range between 4 and 10 \citep[]{Gud2011}.

The seafloor temperature also depends on the surface temperature, $T_{\rm s}$, because we assume that the temperature structure of the ocean is adiabatic (see also \S~\ref{sec:OSM}).
We calculate $T_{\rm s}$ as a function of $P_{\rm CO_2}$, as described in detail  in section~\ref{sec:AM}.
On the area of the seafloor beneath the sorbet region, $T_{\rm floor}$ is equal to the melting temperature at the seafloor pressure.

%===============================================
%% Atmospheric model

\subsection{Atmospheric model} \label{sec:AM}
In this study, we use the open-source code for 1-D radiative-convective climate models, \textit{Atmos}\footnote{https://github.com/VirtualPlanetaryLaboratory/atmos}, developed by Kasting and his collaborators \citep[][]{Kas1993, Kop2013, Ram2014}.
This code calculates radiative fluxes in vertically spacing layers of the atmosphere, using the two-stream approximation with the coefficients for radiative absorption and scattering by gaseous molecules updated by \citet[]{Kop2013}.
We assume a 1-bar N$_2$ atmosphere with various partial pressures of ${\rm CO_2}$.
The distribution of the relative humidity of water vapor is treated according to the empirical Manabe-Wetherald model which assumes the surface relative humidity of 0.8, based on the present Earth's atmosphere \citep[]{Man1967, Pav2000}.
According to \citet[]{Kop2013}, we use the surface albedo of 0.32, which implicitly includes the effects of present-day Earth water clouds.
We use the present insolation flux at the Earth's orbit $S_{\odot}$ (=1360~W~m$^{-2}$) and the present Sun's spectrum as the fiducial value and spectrum model, respectively.
The other model settings are the same as those adopted in \citet[]{Ram2014}.
Then, we calculate equilibrium values of $T_{\rm s}$ as a function of $P_{\rm CO_2}$ for given stellar insolation, using a time-stepping approach with moist convective adjustment \citep[]{Pav2000}.
We have confirmed that our calculated $T_{\rm s}$ is almost the same with sufficient accuracy as that from \citet[]{Ram2014}. 
We discuss the uncertainties and impacts of stellar insolation, 
surface albedo, and relative humidity in sections~\ref{sec:DSL}, \ref{sec:uAM}, and \ref{sec:EXO}.

  %===============================================
  %% Numerical procedure
\subsection{Numerical procedure} \label{sec:putting}

	\begin{table*}
    \begin{center}
      \caption{Variables and their values. \label{Table3}}
      \begin{tabular}{ccc} \hline
        Parameter & Symbol & Value  \\ \hline
        Ocean mass & $M_{\rm oc}$ & 1--200~$M_{\rm oc,\oplus}$  \\
        Mean mantle heat flow & $\bar{q}$ & 40, 60, 80, 100, 120 mW~${\rm m^{-2}}$ \\
        Activation energy of seafloor weathering & $E_{\rm a}$ & 30, 41, 92 kJ~${\rm mol^{-1}}$  \\
        ${\rm CO_2 / H_2O}$ molar ratio & $\gamma$ & $1.0\times10^{-3}$--$10$ \\
        \hline
      \end{tabular}
    \end{center}
  \end{table*}

  \begin{table*}
    \begin{center}
      \caption{Parameters and their values. \label{Table4}}
      \begin{tabular}{ccc} \hline
        Parameter & Symbol & Value  \\ \hline
        Earth ocean mass & $M_{\rm oc, \oplus}$ & $1.37 \times 10^{21}$ kg \\
        Molar quantity of ${\rm H_2O}$ in the Earth ocean mass & $n_{\rm oc,\oplus}$ & $7.6\times 10^{22}$ mol \\
        Thermal conductivity of the HP ice & $k$ & 3.8 W ${\rm m^{-1}}$ ${\rm K^{-1}}$  \\
        Constant for the viscosity of the HP ice & $B$ & $6.7 \times 10^{19}$ ${\rm Pa^{4.5}} {\rm s}$  \\
        Characteristic shear stress of the HP ice & $\zeta$ & $2.0\times 10^6$ Pa\\
        Activation energy for the viscosity of the HP ice & $E^*$ & 110 kJ ${\rm mol^{-1}}$ \\
        Activation volume for the viscosity of the HP ice & $V^*$ & $1.1\times 10^{-5}$ ${\rm m^3}$ ${\rm mol^{-1}}$ \\
        Critical Rayleigh number & $Ra_{\rm cr}$ & 2000 \\
        Thermal conductivity of the oceanic crust & $k_{\rm rock}$ & 3.3 W ${\rm m^2}$ ${\rm s^{-1}}$ \\
        Thermal diffusivity of the oceanic crust & $\kappa_{\rm rock}$ & $1.0\times 10^{-6}$ ${\rm m^2}$ ${\rm s^{-1}}$ \\
        Present Earth's seafloor weathering rate & $F_{SW}^*$ & $2.0\times 10^{12}$ mol ${\rm yr^{-1}}$ \\
        Reference seafloor temperature & $T_0$ & 289 K  \\
        Area of the oceanic floor & $A_0$ & $3.1\times10^{14}$ ${\rm m^2}$\\
        Regassing ratio & $\beta$ & 0.4\\
        \hline
      \end{tabular}
    \end{center}
  \end{table*}

In summary, for given values of ocean mass $M_{\rm oc}$ and mean mantle heat flow $\bar{q}$, we determine the climate of the ocean planet by the following procedure.
\begin{itemize}
	\item[(i)]
	For trial values of surface temperature $T_{\rm s}$ and surface pressure $P_{\rm s}$,
	we integrate Eqs.~(\ref{hy})--(\ref{adiabatic}) inward from the surface to determine temperature as a function of pressure in the ocean (see \S~\ref{sec:OSM}).
	We find a level where the adiabat crosses the melting temperature of ice.
	The layer between the crossover level and the oceanic crust surface consists of HP ice.
	Then, the seafloor pressure $P_\mathrm{floor}$ and the thickness of the HP ice layer $D$ are determined.
	If the adiabat reaches the oceanic crust surface before crossing the ice melting curve, the planet has no ice in the deep ocean.
	The numerical integration is performed with a 4th-order Runge-Kutta method.
	The size of the interval is chosen so that the pressure at the crossover point is determined with $<$~0.1~\% accuracy.
\vspace{0.5\baselineskip}
	\item[(ii)]
	When the HP ice is present, from the seafloor environment model, we determines the critical heat flow  $q_{\rm cr}$ (or the area of the sorbet region) from Eq.~(\ref{conduction}) or (\ref{q_cr}), depending on $Ra$ (\S~\ref{sec:SEM}).
	Then, we obtain the effective weathering area $f_{\rm oc}$ by substituting $q_{\rm cr}$ and ${\bar q}$ in Eq.~(\ref{criticalheat}).
	Also, we obtain the seafloor temperature $T_\mathrm{floor}$ in the sorbet region by substituting $P_\mathrm{floor}$ in Eq.~(\ref{eq:Tmel}).

\vspace{0.5\baselineskip}
	\item[(iii)]
	In the carbon cycle model (\S~\ref{sec:carbon}), using $T_\mathrm{floor}$ and $f_\mathrm{oc}$ obtained above, we perform a time integration of Eqs.~(\ref{c_atoc})--(\ref{c_man}) and determine the carbon partition among the atmosphere, ocean, oceanic crust, and mantle.
 	Then, from the calculated $P_\mathrm{CO_2}$, we obtain a new value of $T_{\rm s}$ (and thereby $P_\mathrm{s}$) from the atmospheric model (\S~\ref{sec:AM}).
	If the new value of $T_\mathrm{s}$ differs by $>$ 0.01~K from the trial value of $T_\mathrm{s}$, we return to Step (i) and repeat the above procedures with the new $T_\mathrm{s}$.
	The time integration is performed with a Euler method and the interval size is chosen so that the time difference in the molar number of carbons is smaller than 0.1 \% for all the reservoirs.

\vspace{0.5\baselineskip}
	\item[(iv)]
	Once all the time derivatives in Eqs.~(\ref{c_atoc})--(\ref{c_man}) become zero, we judge the solution as an equilibrium state.
	If the surface temperature drops below 273 K, we also stop the time integration and regard the solution as a snowball state.

\end{itemize}

We start time-stepping calculations at arbitrarily high $P_{\rm CO_2}$ (i.e., in a warm condition) for finding equilibrium solutions.
We have confirmed that the results are insensitive to choice of the initial condition, provided a sufficiently high CO$_2$ pressure ($P_{\rm CO_2} > 10$~bars) is adopted. 
(The carbon cycle and climate stability in the snowball state are discussed in section~\ref{sec:ucarbon}.) 
In most of our simulations, response against perturbations for the carbon budget in the atmosphere-ocean system is mainly controlled by regassing, the timescale of which is $\sim \tau / \beta = 250$~Myr for ${\bar q} = 80$~mW~m$^{-2}$.
Thus, an equilibrium state is achieved on a timescale of the order of Gyr, which
is also consistent with results shown in \citet[]{Fol2015}.

The parameters and constants with their values adopted in this study are summarized in Tables~\ref{Table3} and \ref{Table4}, respectively.
The upper limit for ocean mass $M_\mathrm{oc}$ that we consider is 200~$M_\mathrm{oc, \oplus}$.
The reasoning is as follows:
We suppose that plate tectonics is working on the planet.
Although still not fully understood, an increase in water has negative effects on plate tectonics.
In particular, it leads to reducing crustal production and degassing, since the solidus temperature of the mantle material increases with pressure \citep[]{Kit2009, Noa2016}.
According to \citet[]{Noa2016}, crustal production completely ceases for an Earth-mass planet with the ocean layer thicker than approximately 400~km, if plate tectonics operates.
The ocean mass of 200~$M_{\rm oc, \oplus}$ that we adopt here corresponds to the ocean layer of $\sim$350~km for $T_{\rm s} = 300$~K.
We do not consider ocean planets with more massive oceans because such planets are expected to have no geochemical cycle.
For planetary climates with no geochemical cycle, see \citet[]{Kit2015} and \citet[]{Kit2018}.

Note that we assume a spherically symmetric structure in the internal structure modeling, while we consider the presence of the sorbet and HP ice regions in the deep ocean in the seafloor environment modeling.
Such self-contradiction, however, has little influence on our whole modeling.
This is because only the thermal structure above the HP ice layer is of interest in this study and the equations of state of water, rock, and iron are rather insensitive to temperature.

  %*************************************** Methods **********************************************

% % %*************************************** Methods **********************************************

% % %*************************************** Results **********************************************

% % %*************************************** Results1 **********************************************

%*************************************** Results1 **********************************************
\section{Melting of the HP ice} \label{sec:melting}

\begin{figure*}
  \includegraphics[width=17cm]{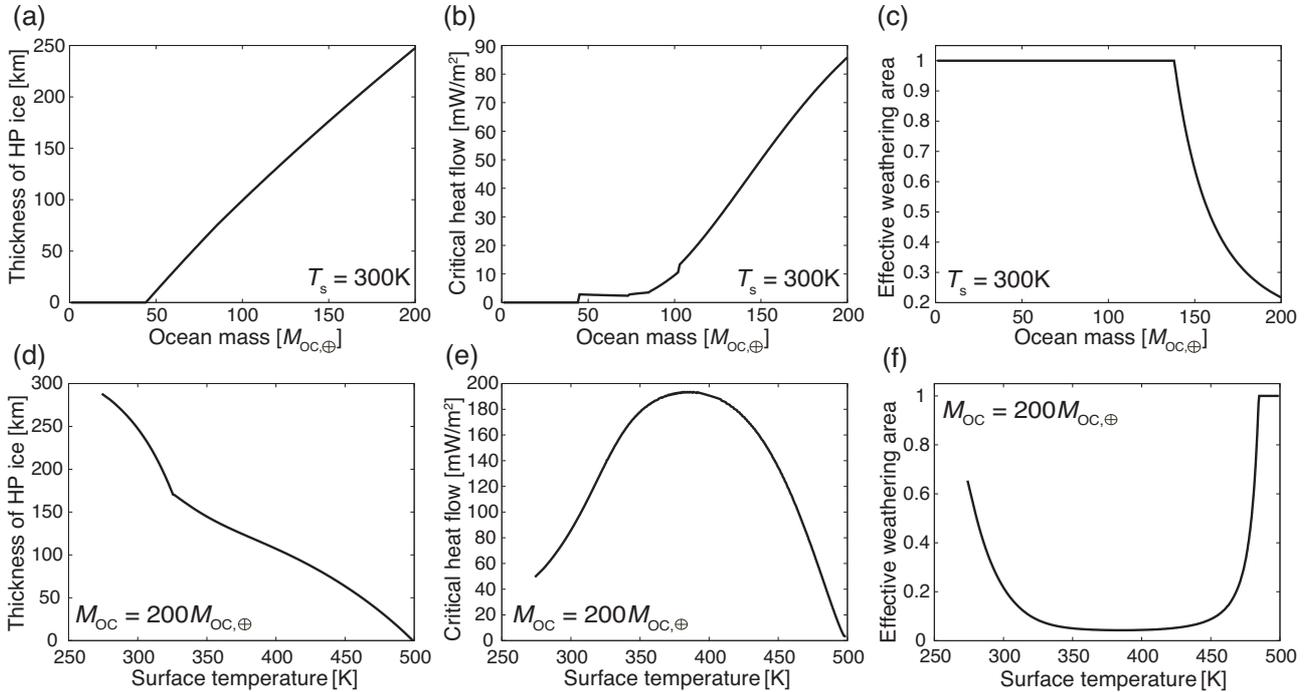}
  \caption{Formation of high-pressure (HP) ice and its impacts on the seafloor condition.
  Thickness of the HP ice layer (panels $a$ and $d$),
  critical heat flow (panels $b$ and $e$), and
  effective weathering area (panels $c$ and $f$) are shown as a function of ocean mass $M_{\rm oc}$ for surface temperature $T_{\rm s} = 300$~K (top)
  and as a function of $T_{\rm s}$ for $M_{\rm oc} = 200 M_{\rm oc, \oplus}$ (bottom).
  ${M_{\rm oc, \oplus}}$ represents the present Earth's ocean mass.
  Note that we have not used the carbon cycle model for determining $T_\mathrm{s}$ here, but performed calculations for given values of $T_\mathrm{s}$, instead.
  \label{Fig2}}
\end{figure*}

We first investigate the behavior of the HP ice with a focus on the effective weathering area, which is a controlling factor for seafloor weathering. Here we do not use the carbon cycle model, but, instead, perform calculations for fixed values of the surface temperature $T_{\rm s}$.
Figure~\ref{Fig2} shows the calculated thickness of the HP ice layer $D$ (left column), the critical heat flow $q_{\rm cr}$ (middle column) and effective weathering area $f_{\rm oc}$ (right column) as a function of ocean mass $M_{\rm oc}$ for $T_{\rm s} = 300$~K (top) and as a function of $T_{\rm s}$ for $M_{\rm oc} = 200M_{\rm oc, \oplus}$ (bottom).
In those calculations, the mean mantle heat flow ${\bar q}$ is assumed to be $80$~mW~m$^{-2}$.

\subsection{Dependence on Ocean Mass}

The overall dependence on ocean mass is as follows.
As shown in Fig.~\ref{Fig2}$a$, the HP ice is present, if $M_{\rm oc} \gtrsim 45M_{\rm oc, \oplus}$.
Its thickness increases almost linearly with ocean mass and reaches 247~km at $M_\mathrm{oc}$ = $200M_{\rm oc, \oplus}$.
In Fig.~\ref{Fig2}$b$, the critical heat flow is found to be zero for $M_{\rm oc} \lesssim 45 M_{\rm oc, \oplus}$, because of no HP ice, and then increase with ocean mass, up to about $80$~mW~m$^{-2}$ ($\simeq {\bar q}$) at $M_{\rm oc} = 200 M_{\rm oc, \oplus}$.
In Fig.~\ref{Fig2}$c$, the effective weathering area is found to be unity until $M_{\rm oc} \simeq 139 M_{\rm oc, \oplus}$ and rapidly decrease to about 0.2 at $M_{\rm oc} = 200 M_{\rm oc, \oplus}$.

A jump in $q_{\rm cr}$ is found at $M_\mathrm{oc} \simeq 74 M_{{\rm oc},\oplus}$ in Fig.~\ref{Fig2}$b$.
At that point, the heat transport mechanism in the HP ice above the critical point (i.e., $q=q_{\rm cr}$) changes from conduction to convection.
For $M_{\rm oc} \lesssim 74 M_{\rm oc, \oplus}$ ($D \simeq$ 55~km), the HP ice layer is thin enough and, therefore, the temperature difference $\Delta T_{\rm HP}$ (= $T_{\rm BB}^{\rm mel} - T_{\rm TB}^{\rm mel}$) is small enough for conduction to transport the heat flux from the oceanic crust.
However, as shown in Fig~\ref{Fig2}$c$, $q_\mathrm{cr}$ $\simeq$ 10~mW~m$^{-2}$ $< \bar{q}/2$ at $M_{\rm oc} \lesssim 74 M_{\rm oc, \oplus}$, meaning that the HP ice is entirely molten (i.e., $f_{\rm oc} = 1$), that is, the seafloor is covered entirely with the sorbet for $M_{\rm oc} \lesssim 74 M_{\rm oc, \oplus}$ (see the text just below Eq.~[\ref{criticalheat}]).
Note that discontinuities in $q_{\rm cr}$ or $dq_{\rm cr}/dM_{\rm oc}$ found at $M_{\rm oc} \simeq 86$ and $ 103 M_{\rm oc, \oplus}$ come from those in the melting curve of ${\rm H_2O}$ at the phase boundaries of ice~\rom{6}/\rom{7}.

The critical heat flow exceeds $\bar{q}/2$ at $M_{\rm oc} \simeq$ 139~$M_{\rm oc, \oplus}$ ($D \simeq 160$ km), until which the effective weathering area is unity, and then increases further with ocean mass.
Such an increase in $q_{\rm cr}$ occurs because the Rayleigh number in the HP ice layer increases.
Thus, the effective weathering area decreases with ocean mass, but never becomes zero until $M_\mathrm{oc} = 200M_{\rm oc, \oplus}$.
This means that water-rock reactions between water and rock including the seafloor weathering are possible, despite the presence of the thick HP ice, because the sorbet region also exists near the mid-ocean ridge.

%%%%%%%%%%%%%%%%%%%%%%%%%%%%%%%%%%%%%%
\subsection{Dependence on Surface Temperature} \label{sec:DST}

\begin{figure}
  \includegraphics[width=\columnwidth]{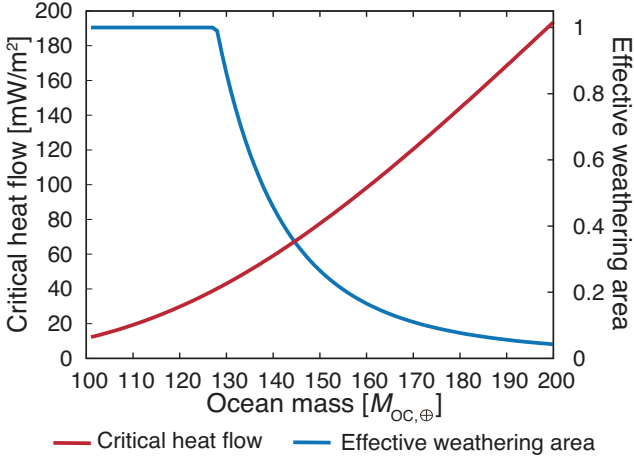}
  \caption{
  The maximum of critical heat flow (red solid line) and the minimum of effective weathering area (blue solid line) found in the surface temperature range considered in this study are shown as a function of ocean mass.
  In these calculations, we have assumed the mean mantle heat flux ${\bar q} = 80$~mW~m$^{-2}$.
  \label{Fig3}}
\end{figure}

The three lower panels of Fig.~\ref{Fig2} show the dependence on the surface temperature for $M_{\rm oc}$ = $200 M_{\rm oc, \oplus}$.
The HP ice thickness decreases, as the surface temperature increases, as shown in Fig.~\ref{Fig2}$d$.
At $T_{\rm s} \simeq 320$~K, the curve is a bit inflected.
This is due to the phase change of HP ice from ice~\rom{6} to ice~\rom{7}.

In Figs.~\ref{Fig2}$e$ and \ref{Fig2}$f$, we find a maximum of the critical heat flow and a minimum of the effective weathering area, respectively, at $T_\mathrm{s} \simeq 390$~K.
As indicated in Eq.~(\ref{q_cr}), $q_{\rm cr}$ depends on $\eta_{\rm BBL}$ and $\delta$, both of which decrease with $T_{\rm s}$.
For $T_{\rm s} \lesssim$ 390~K, $\delta^4$ decreases more rapidly than $\eta_{\rm BBL}$ and, thus, $q_{\rm cr}$ increases with $T_{\rm s}$. In contrast, for $T_{\rm s} \gtrsim$ 390~K, the latter dominates over the former, so that $q_{\rm cr}$ decreases.
At $T_{\rm s} \simeq$ 390~K, $\partial (\eta_{\rm BBL} / \delta^{4})/\partial T_{\rm s} = 0$.
The behavior of the effective weathering area can be readily understood from Eq.~(\ref{criticalheat}), namely, $f_{\rm oc} \propto q_{\rm cr}^{-2}$.
The minimum is $f_{\rm oc} \simeq$ 0.04.
% \ikome{Eq.~[\ref{criticalheat}]を引用するだけでいいと思う→}
% \sakujo{
% Also, effective weathering area increases sharply with increasing the surface temperature, in particular for $T_\mathrm{s} \gtrsim 450$~K.
% When $T_{\rm s} \lesssim 390$~K, although the thickness and temperature difference of the HP ice increase with decreasing $T_\mathrm{s}$, the critical heat flow decreases with decreasing the surface temperature.
% Lower surface temperature results in a cooler BBL with high viscosity.
% High viscosity leads to a thick BBL for a given heat flux, which results in the smaller temperature gradient in BBL.
% Thus, effective weathering area increases with decreasing the surface temperature.
% Although the bottom temperature of the HP ice layer varies with the ocean mass, the behavior above is unchanged for any ocean mass.}

In Fig.~\ref{Fig3}, we show the maximum of critical heat flow $q_{\rm cr, max}$ and minimum of effective weathering area $f_{\rm oc, min}$ as a function of ocean mass for ${\bar q} = 80$~mW~m$^{-2}$.
Here we show only the results for the case of convective HP ice for $M_{\rm oc} > 100M_{\rm oc, \oplus}$ because the critical heat flow due to conduction is small.
While $q_{\rm cr, max}$ is found to monotonically increase with $M_{\rm oc}$,
$f_{\rm oc, min}$ begins to drop with $M_{\rm oc}$ at $M_{\rm oc} \simeq 128 M_{\rm oc, \oplus}$, which is smaller than in the case of $T_{\rm s} = 300$~K because of difference in $T_{\rm s}$.
The blue line in Fig.~\ref{Fig3} indicates that even the minimum of $f_{\rm oc}$ is unity for $M_{\rm oc} \lesssim 128 M_{\rm oc, \oplus}$, which means that the HP ice is entirely molten and the seafloor is completely covered with the sorbet, regardless of surface temperature, in such an ocean mass range for the Earth-like mean mantle heat flow (${\bar q} = 80$~mW~m$^{-2}$).
Also, $f_{\rm oc, min} > 0$, meaning that seafloor weathering works, even if $M_{\rm oc} = 200M_{\rm oc, \oplus}$.

%*************************************** Results1 **********************************************

% % %*************************************** Results2 **********************************************

%*************************************** Results2 **********************************************
%===============================================
%% Seafloor weathering enhanced by the HP ice
\section{Seafloor weathering enhanced by the HP ice} \label{sec:ESW}

\begin{figure*}
  \includegraphics[width=17.5cm]{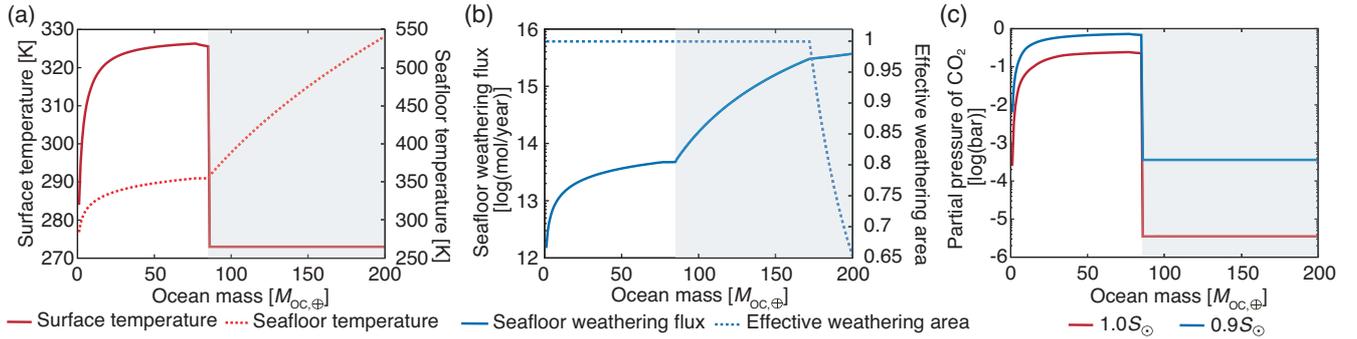}
  \caption{Surface and seafloor conditions obtained from the carbon cycle model:
  ($a$) Surface temperature (red sold line) and seafloor temperature (red dashed line), ($b$) seafloor weathering flux (blue solid line) and effective weathering area (blue dashed line), and ($c$) partial pressure of ${\rm CO_2}$ are shown as a function of ocean mass in the unit of the Earth's ocean mass $M_{\rm oc, \oplus}$.
  Shaded is the range for the snowball state.
  In this calculation, the mean mantle heat flow is assumed to be 80~mW~m$^{-2}$.
  The symbol $S_{\odot}$ represents the solar insolation received by the present Earth.
  \label{Fig4}}
\end{figure*}

\subsection{Consequence of carbon cycle}
Here we examine the planetary climate based on the carbon cycle including the effective weathering area obtained above.
The calculation results for ${\bar q} = 80$~mW~m$^{-2}$, $\gamma = 0.19$, and $S=S_\odot$ and $0.9 S_\odot$ are shown in Fig.~\ref{Fig4},
where ($a$) the surface and seafloor temperatures, ($b$) the seafloor weathering flux and effective weathering area, and ($c$) the partial pressure of atmospheric ${\rm CO_2}$ are plotted as functions of the ocean mass.
In Fig.~\ref{Fig4}$a$, two obviously different states are found: 
One is the state with $T_{\rm s} > 273$~K, where the carbon cycle is in a steady state, the other, as indicated by a shaded area, is the state with $T_{\rm s} = 273$~K, where the carbon cycle calculation is artificially stopped at $T_{\rm s} = 273$~K because the surface ice is expected to form (see also \S~\ref{sec:putting}).
The former is called the \textit{equilibrium state} and the latter is called the \textit{snowball state} in this study. 
In this case, the HP ice begins to form at $M_\mathrm{oc}$ = $86 M_{\rm oc, \oplus}$.
It turns out that the formation of the HP ice has a drastic effect on the carbon cycle and determines which state is achieved.

In the case of no HP ice (i.e., $M_{\rm oc} < 86 M_{\rm oc, \oplus}$), both the surface temperature and CO$_2$ partial pressure increase with ocean mass.
An equilibrium state is achieved for a given ocean mass via a negative feedback loop such that an increase in $P_{\rm CO_2}$ raises the surface temperature, which leads to a rise in seafloor temperature, which enhances seafloor weathering flux, which finally reduces the atmospheric CO$_2$.
The larger the ocean mass, the larger the total carbon inventory $C_{\rm total}$ is (see Eq.~[\ref{Ctotal}]).
Since an increase in $C_{\rm total}$ enhances the degassing flux of CO$_2$ (see Eq.~[\ref{degassing}]), the surface temperature consequently raises with ocean mass.
This is, in other words, because the enhancement of the degassing flux dominates over the increase in seafloor temperature in the case of $\gamma = 0.19$.
While the outcome depends on $\gamma$,
we have confirmed that this trend is the same also in the case of one-tenth of the Earth-like value $\gamma$ (= 0.019) and comet-like $\gamma$ (= 0.71) higher than the Earth's:
The equilibrium values of $T_{\rm s}$ and $P_{\rm CO_2}$ for $M_{\rm oc} < 86M_{\rm oc, \oplus}$ are increased up to 326~K and $2.4 \times 10^{-1}$~bars for $\gamma = 0.19$ and $S_{\rm \odot}$, respectively (see Figs.~\ref{Fig4}$a$ and \ref{Fig4}$c$).

In contrast, when the HP ice is present ($M_{\rm oc} \geq 86 M_{\rm oc, \oplus}$), the negative feedback never works and, consequently, the snowball state is attained.
This is because the seafloor temperature on the area under the sorbet region, where seafloor weathering works, is fixed at the melting temperature of ice and, thus, insensitive to the surface temperature.
Although the reduction in effective weathering area reduces seafloor weathering rate (see Fig.\ref{Fig4}$b$), it is found to have little impact on surface temperature because the seafloor weathering flux is significantly higher than the degassing flux.

\subsection{Dependence on stellar insolation} \label{sec:DSL}
We examine the dependence of planetary climate on stellar insolation.
Since the runaway greenhouse limit, which controls the inner edge of the habitable zone, is only slightly higher than $S_\odot$ \citep[e.g., $1.06S_{\odot}$][]{Kop2013}, we show only the results for smaller stellar insolation of $0.9S_\odot$ than the fiducial value of $1.0S_{\odot}$.
As shown in Fig.~\ref{Fig4}$c$, stellar insolation affects CO$_2$ partial pressure both in the equilibrium and snowball states: the smaller the stellar insolation, the higher the CO$_2$ pressure is, as a whole:
$P_\mathrm{CO_2}$ for $S = 0.9 S_\odot$ is higher by a factor of $\sim$ 3 and by two orders of magnitude than that for $S = 1.0 S_\odot$ in the equilibrium and snowball states, respectively. 
The other quantities are almost unaffected by stellar insolation.
This is because the increase in CO$_2$ pressure compensates for the decline in stellar insolation so as not to change the surface temperature which controls weathering behavior and \comhp\, in our climate model.
Thus, variation in stellar insolation has little impact on planetary climate, provided the planet is located in the habitable zone.
%Further discussion about is given in \S~\ref{sec:uAM}.

\subsection{Dependence on mean mantle heat flow}

\begin{figure}%[H]
  \begin{center}
    \includegraphics[width=7.0cm]{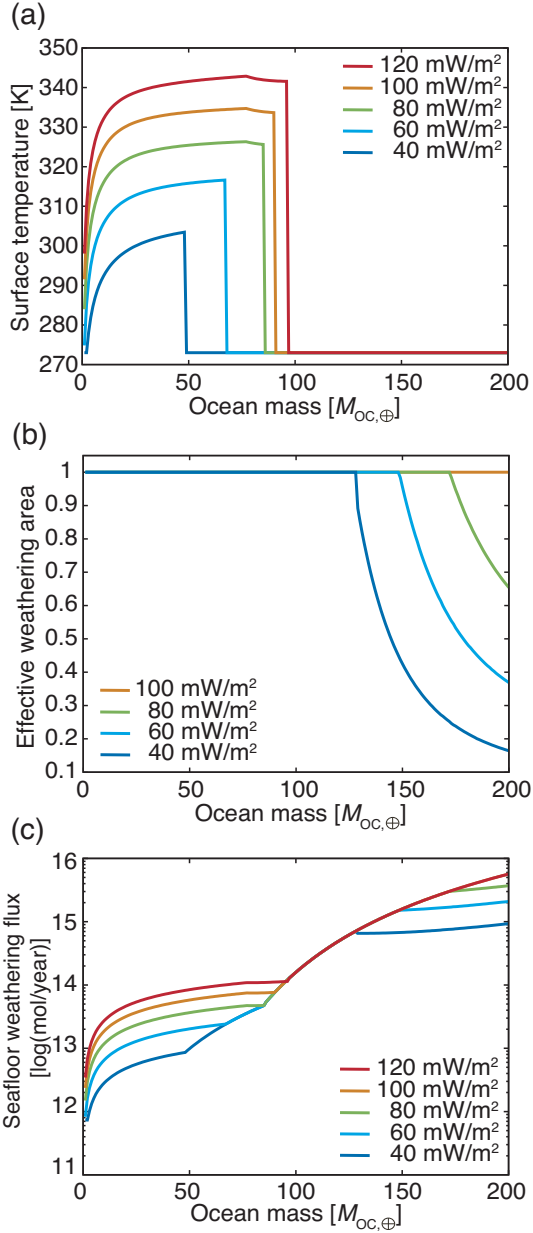}
    \caption{Surface and seafloor conditions obtained from the carbon cycle model for five different values of the mean mantle heat flow.
    ($a$) Surface temperature, ($b$) effective weathering area and ($c$) seafloor weathering flux are shown as functions of ocean mass.
    \label{Fig5}}
  \end{center}
\end{figure}

Next, we examine what impact the mean mantle heat flow $\bar{q}$ has on the surface and seafloor conditions.
Figure~\ref{Fig5} shows ($a$) the surface temperature, ($b$) effective weathering area, and ($c$) seafloor weathering flux for five different choices of $\bar{q}$.
The variation in mean mantle heat flow turns out to yield no change on the overall behavior, but quantitative modifications to the ocean mass dependence.

First, as seen in Fig.~\ref{Fig5}$a$, when no HP ice is present, the larger the mean mantle heat flow, the higher the surface temperature is for a given ocean mass.
The variation in $\bar{q}$ leads to a large difference in the surface temperature (up to  40~K).
Also, the surface condition lapses into the snowball state at larger ocean mass for a larger $\bar{q}$.
That is because as $\bar{q}$ increases, the seafloor spreading rate $A_{\rm S}$ increases (see Eq.~[\ref{SR}]) and, thus, the degassing flux increases (see Eq.~[\ref{degassing}]), leading to higher surface temperature and larger critical ocean mass for forming the HP ice (\comhp, see also Fig.~\ref{Figa1}).
In Fig.~\ref{Fig5}$c$, the seafloor weathering flux is also found to increase by approximately an order of magnitude in response to the rise in the degassing flux.

As shown in Fig.~\ref{Fig5}$b$, the effective weathering area $f_{\rm oc}$ starts to decrease from unity at larger ocean mass for larger mean mantle heat flow and is always unity until $M_{\rm oc} = 200 M_{\rm oc, \oplus}$ for ${\bar q} \geq 100$~mW~m$^{-2}$.
While the seafloor weathering flux changes with $\bar{q}$ (i.e., $f_{\rm oc}$) in the case with HP ice, the reduction in $f_{\rm oc}$ turns out to have only a small effect on the surface temperature, because of significantly high seafloor weathering rate for any value of $\bar{q}$ (see Fig.~\ref{Fig5}$c$).

\begin{figure*}
  \includegraphics[width=17cm]{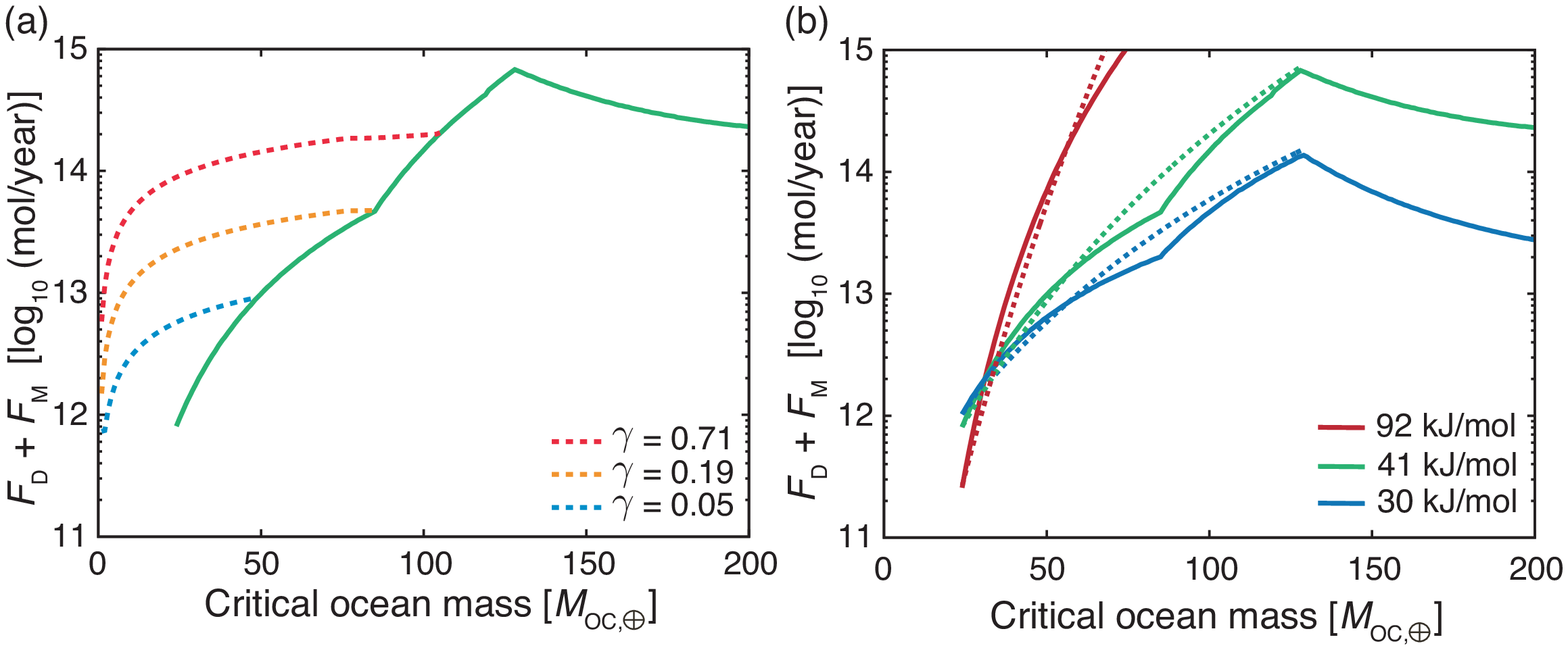}
  \caption{Relationship between the critical ocean mass for the snowball state (\comsb) and the total-degassing flux $F_{\rm D}+F_{\rm M}$ (solid lines) for different choices of the $\rm CO_2/H_2O$ molar ratio in the source of volatiles, $\gamma$ (see Eq.~[\ref{Ctotal}]), and seafloor weathering activation energy, $E_{\rm a}$ (see Eq.~[\ref{SW}]).
  Panel~($a$) shows the fiducial case with $E_{\rm a} = 41$~kJ~mol$^{-1}$.
  For reference, dashed lines represent the relationship between $F_{\rm D} + F_{\rm M}$ and not the \comsb, but just the ocean mass for three different values of $\gamma$ when the planetary climate is in an equilibrium state.
  Panel~($b$) compares the results for three different values of $E_{\rm a}$.
  Dashed lines represent the analytical solutions of \comsb\ (see Eq.~[\ref{eq:moc_cr}]).
  On the right side of the solid lines, the sorbet is present in the deep ocean.
  In these calculations, we have assumed the mean mantle heat flux ${\bar q} = 80$~mW~m$^{-2}$.
  \label{Fig6}}
\end{figure*}

\subsection{Dependence on $\rm CO_2/H_2O$ ratio and seafloor weathering activation energy}
As described in section~\ref{sec:carbon}, the carbon cycle depends on the total carbon inventory and seafloor weathering rate.
The former may differ greatly from planet to planet, as suggested, for example, by a difference in the $\rm CO_2/H_2O$ molar ratio $\gamma$ (Eq.~[\ref{Ctotal}]) between comets and the Earth.
Also, the seafloor weathering rate is in general uncertain, mainly because the activation energy $E_{\rm a}$ (Eq.~[\ref{SW}]) is poorly determined observationally.
Here we investigate the sensitivities of the planetary climate to $\gamma$ and $E_{\rm a}$ with focus on
the critical ocean mass, beyond which the planetary climate is in the snowball state (hereafter, abbreviated to \comsb\, and denoted by $\Mcrsb$).

In Fig.~\ref{Fig6},
we plot the relationships between $\Mcrsb$ and total degassing flux, $F_{\rm D} + F_{\rm M}$, for various values of $\gamma$ between $7.4 \times 10^{-3}$ and 2.1; both $\Mcrsb$ and $F_{\rm D}+F_{\rm M}$ are obtained from the carbon cycle calculations.
Here we assume ${\bar q} = 80$~mW~m$^{-2}$.
Fig.~\ref{Fig6}$a$ shows the fiducial case with $E_{\rm a} = 41$~kJ~mol$^{-1}$;
Fig.~\ref{Fig6}$b$ shows cases with three different values of $E_{\rm a}$.
For reference, in Fig.~\ref{Fig6}$a$, we show the relationships between $F_{\rm D}+F_{\rm M}$ and not $\Mcrsb$ but $M_{\rm oc}$ for three different values of $\gamma$ by dashed lines (the result for $\gamma = 0.19$ is also shown in Fig.~\ref{Fig4}$b$).
% \sakujo{, which reaches the curve of $\Mcrsb$ at $M_{\rm oc} = 86 M_{\rm oc, \oplus}$}.
In Fig.~\ref{Fig6}$a$, we can see that the total degassing flux increase almost linearly with $\gamma$ for a given ocean mass.
For $\gamma < 7.4 \times 10^{-3}$, the \comsb\, is absent because the snowball state is achieved in all the ocean mass range due to low degassing flux. %\sakujo{\uwave{because small degassing flux does not make warm equilibrium climate ($T_{\rm s} > 273$~K)}}
% \sakujo{For $\gamma > 1.1$, we do not present the calculation results, because our atmospheric model is inapplicable for such high surface temperatures ($T_{\rm s} > 380$~K, see Appendix~\ref{sec:AM}).}

As shown in Fig.~\ref{Fig6}$a$, the total degassing flux has a peak at $\Mcrsb = 128 M_{\rm oc, \oplus}$.
For $\Mcrsb \leq 128 M_{\rm oc, \oplus}$, the \comsb\, increases from 24 to 128~$M_{\rm oc, \oplus}$ and the total degassing flux, which is determined by $F_{\rm SW} (T_{\rm floor}, f_{\rm oc}) = F_{\rm SW} (T^{\rm mel}_{\rm floor}, 1)$, increases from $8.1~\times~10^{11}$ to $6.8~\times~10^{14}$~mol~yr$^{-1}$ with increase in $\gamma$ from $7.4 \times 10^{-3}$ to $2.1$.
Despite order-of-magnitude variation in $F_{\rm D}+F_{\rm M}$, the \comsb\, varies moderately by a factor of $\sim$5 (see section~\ref{sec:app} for an analytical interpretation).
On the other hand, for $\Mcrsb > 128 M_{\rm oc, \oplus}$, the total degassing flux is determined by the minimum weathering flux with the HP ice, namely $F_{\rm SW} (T_{\rm floor}, f_{\rm oc}) = F_{\rm SW} (T^{\rm mel}_{\rm floor}, f_{\rm oc, min})$ (see Fig.~\ref{Fig3} for $f_{\rm oc, min}$).
Thus, the \comsb\, increases from 128 to $200M_{\rm oc, \oplus}$ and the total degassing flux decreases from $6.8~\times~10^{14}$ to $2.3~\times~10^{14}$~mol~yr$^{-1}$ with decrease in $\gamma$ from $2.1$ to $5.2 \times 10^{-1}$.
In this diagram, equilibrium climates ($F_\mathrm{SW}$ = $F_\mathrm{D}+F_\mathrm{M}$) are achieved on the side above the solid line, whereas the planetary surface condition lapses into snowball states ($F_\mathrm{SW} > F_\mathrm{D}+F_\mathrm{M}$), because of the presence of the sorbet region, on the side below the solid line.
Note the curve of $\Mcrsb$ is a bit inflected at $M_{\rm oc} = 85M_{\rm oc, \oplus}$ because of a phase change of the HP ice.

In Fig.~\ref{Fig6}$b$, we show the impact of $E_{\rm a}$ on $\Mcrsb$.
The curve for larger $E_{\rm a}$ is found to be steeper.
The three curves cross each other at $F_{\rm D}+F_{\rm M} = 2.0 \times 10^{12}$~mol~yr$^{-1}$,
where the seafloor temperature $T_{\rm floor}$ is equal to $T_0$ (=$289$~K), so that $F_{\rm SW}$ is independent of $E_{\rm a}$ (see Eq.~[\ref{SW}]).
Above the crossover point, the higher the activation energy, the smaller the \comsb\, is;
its dependence is opposite below the crossover point.
Although being large relative to that on $\gamma$, the dependence of $\Mcrsb$ on $E_{\rm a}$ is at most linear.
Thus, it would be fair to say that $\Mcrsb$ is rather insensitive to $E_{\rm a}$.
We further discuss the nature of the \comsb\, analytically in section~\ref{sec:app}.

%*************************************** Results2 **********************************************

% % %*************************************** Results **********************************************

% % %*************************************** Discussion **********************************************

%*************************************** Discussion **********************************************
\section{Discussion}

\subsection{Critical ocean mass for snowball state}  \label{sec:app}
One of the most important findings in this study is that there is a critical ocean mass, beyond which an ocean terrestrial planet has an extremely cold climate (i.e., the snowball state).
Furthermore, we have found that the \comsb, $\Mcrsb$, falls into a relatively narrow range between 20 and 100~$M_{\rm oc, \oplus}$.
Here we give an interpretation to the low sensitivity of $\Mcrsb$ to
the planetary mass $M_\mathrm{p}$, the total degassing flux $F_{\rm D}+F_{\rm M}$, and the activation energy of seafloor weathering $E_{\rm a}$, by deriving an approximate solution for $\Mcrsb$.
This could help us obtain an integrated view of planetary climate on ocean planets under our idealized seafloor environments.

As demonstrated in section~\ref{sec:ESW}, the planetary climate lapses into the extremely cold one, when HP ice is formed on the seafloor.
Then, the seafloor temperature $T_{\rm floor}$ is fixed at the melting temperature $T^\mathrm{mel}$ and thus determined uniquely by the seafloor pressure $P_\mathrm{floor}$.
Since the ocean mass and depth are negligibly small relative to the planetary mass and radius, respectively, under
hydrostatic equilibrium, the seafloor pressure is given approximately by
\begin{eqnarray}
	P_{\rm floor}
	&\approx&
		\frac{GM_{\rm p}}{4 \pi R_{\rm p}^4} M_{\rm oc} =
		\frac{G {\bar \rho_{\rm p}}}{3R_{{\rm p}}} M_{\rm oc},
		\label{eq:hy}
\end{eqnarray}
where $R_{\rm p}$ and ${\bar \rho_{\rm p}}$ are the planetary radius and mean density, respectively.
For $M_{\rm oc}$ = $\Mcrsb$, $P_\mathrm{floor}$ corresponds to the crossover pressure between the adiabat and the melting curve, both of which are independent of planetary mass.
Thus, from Eq.~(\ref{eq:hy}), $\Mcrsb \propto$ $R_{\rm p}^4 / M_{\rm p}$.
According to \citet{Val2007b}, the mass-radius relationship for Earth-like planets
is $R_{\rm p} \propto M_{\rm p}^{0.262}$, which yields $\Mcrsb \propto M_{\rm p}^{0.048}$.
This indicates that the \comsb\, is insensitive to planetary mass; indeed,
between $M_{\rm p}$ = $1M_{\oplus}$ and $10M_{\oplus}$, for example, $\Mcrsb$ differs only by $\sim$12\%.

To derive the dependence of $\Mcrsb$ on $F_{\rm D}+F_{\rm M}$ and $E_{\rm a}$, we consider seafloor weathering.
In the equilibrium state, since $F_\mathrm{SW} (T_\mathrm{floor})$ = $F_\mathrm{D}+F_\mathrm{M}$, $T_{\rm floor}$ is given as a function of $F_{\rm D}+F_{\rm M}$ (see Eq.~[\ref{SW}]).
Also, $T_{\rm floor}$ = $T^{\rm mel}$, when $M_{\rm oc}$ = $\Mcrsb$: From Eq.~(\ref{eq:Tmel}),
\begin{equation}
	T^{\rm mel} \simeq c_1 + c_2 P_{\rm floor},
	\label{eq:fitting}
\end{equation}
where $c_1 = 236$ K, $c_2 = 6.09 \times 10^{-8}$~K~Pa$^{-1}$.
From Eqs.~(\ref{SW})--(\ref{eq:fitting}), $\Mcrsb$ is expressed as
\begin{eqnarray}
	M^{\rm cr}_{\rm oc} =
	\frac{3 R_{\rm p}}{c_2 G {\bar \rho_{\rm p}}}
	\left\{\frac{T_0}{1- \frac{R T_0}{E_{\rm a}} {\rm ln} \left( \frac{F_{\rm D}+F_{\rm M}}{f_{\rm oc} F_{\rm SW}^*}\right) } - c_1 \right\} \label{eq:moc_cr}.
\end{eqnarray}
This equation confirms that the \comsb\, depends on the degassing flux only weakly.
Also, since the denominator of the first term must be positive, the sensitivity of $\Mcrsb$ to $E_{\rm a}$ turns out to be small.
In Fig.~\ref{Fig6}$b$, we plot the relationships between $\Mcrsb$ and $F_{\rm D}+F_{\rm M}$ calculated from Eq.~(\ref{eq:moc_cr}), which is found to reproduce the numerical results well, except for the effect of phase change of the HP ice.

%===============================================
\subsection{Effect of supply limit of cations \label{sec:DPC}}

\begin{figure}
	\includegraphics[width=\columnwidth]{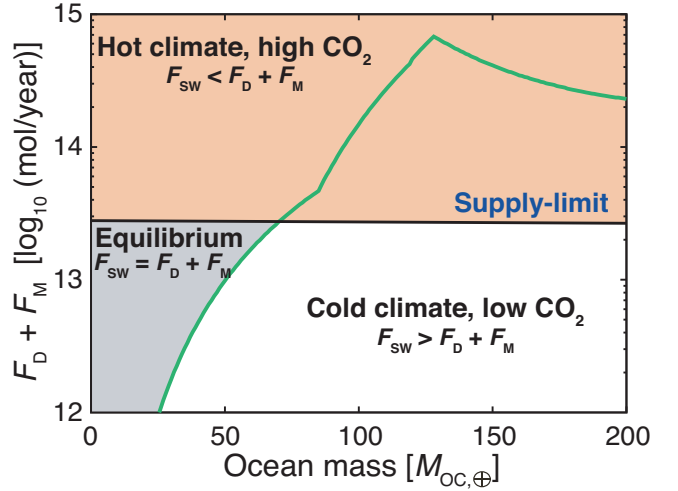}
	\caption{Climate diagram that shows different climate regimes on the plane of the total degassing flux ($F_{\rm D}+F_{\rm M}$) vs. ocean mass.
	The horizontal black solid line represents the supply limit above which seafloor weathering $F_{\rm SW}$ is limited by the insufficient supply of cations (see Eq.~[\ref{supply-limit}]).
	The green solid line corresponds to the critical ocean mass (\comsb), namely, the boundary between the equilibrium states and the extremely cold (snowball) states that would be achieved if no supply limit were assumed, same as that in Fig.~\ref{Fig6}~$a$.
% 	\sakujo{The solid part of the green line is the same as that in Fig.~\ref{Fig6}$a$, whereas the dashed part represents extrapolated results (see the text for the details).}
	In this calculation, we have assumed the mean mantle heat flux ${\bar q} = 80$~mW~m$^{-2}$ and activation energy $E_{\rm a} = 41$~kJ~mol$^{-1}$.
	\label{Fig7}}
\end{figure}

As shown in section~\ref{sec:ESW}, without any limit to seafloor weathering, the presence of the HP ice (exactly to say, the sorbet) always enhances seafloor weathering, resulting in extremely cold climates (i.e., the snowball states).
In reality, however, the seafloor weathering rate is limited by the number of cations available in the oceanic crust.
This is because seafloor weathering occurs through hydrothermal circulation in the oceanic crust and thus the amount of cations available depends on the depth of hydrothermal circulation.
This limit to seafloor weathering rate, which we call the supply limit, $F^{\rm limit}_{\rm SW}$, can be given by \citep[]{Sle+2001, Fol2015}
\begin{equation}
	F^{\rm limit}_{\rm SW} = \frac{x}{m_{\rm c}} \rho_{\rm rock} d_{\rm hy} A_{\rm S} , \label{supply-limit}
\end{equation}
where $x$ is the number fraction of cations (${\rm Ca}^{2+}$, $\mathrm{Mg}^{2+}$, and $\mathrm{Fe}^{2+}$) in the oceanic crust, $m_{\rm c}$ is the averaged molar mass of cation, $\rho_{\rm rock}$ is the density of the oceanic crust, $d_{\rm hy}$ is the depth at which hydrothermal carbonation occurs, and $A_{\rm S}$ is the seafloor spreading rate.
In the present Earth condition, where $x = 0.3$, $m_{\rm c} = 55$~g~mol$^{-1}$, $\rho_{\rm rock} = 2800$~kg~m$^{-3}$ and $d_{\rm hy} = 500$~m \citep[]{Sle+2001}, $F^{\rm limit}_{\rm SW} = 7.6 \times 10^6 A_{\rm S}$.

If the total degassing flux is higher than the supply limit, the atmospheric ${\rm CO_2}$ continues to increase with age.
Qualitatively, the more the atmospheric ${\rm CO_2}$, the higher the surface temperature is.
While climate sensitivity to the amount of ${\rm CO_2}$ is unclear for high $P_{\rm CO_2}$ because of poor understanding of radiative forcing of water vapor for hot atmospheres, recent 1-D radiative-convective calculations show that $T_{\rm s} > 350$~K for $P_{\rm CO_2}$ $>$ several bars, provided the stellar insolation is equal to that for the present Earth \citep[]{Wor2013b, Ram2014}.

Figure~\ref{Fig7} is the climate diagram for $\bar{q} = 80$~mW~m$^{-2}$ and $E_{\rm a} = 41$~kJ~mol$^{-1}$, where we indicate three different climate regimes, which include the equilibrium climates, the extremely cold climates (or the snowball states), and the extremely hot climates.
The extremely hot climate is a state such that $F_\mathrm{SW}$ $<$ $F_\mathrm{D}+F_\mathrm{M}$, because of supply limit so that CO$_2$ accumulates in the atmosphere.
The supply limit (horizontal black solid line) is calculated from Eq.~(\ref{supply-limit}).
The boundary between the equilibrium-climate and cold-climate regimes (green line) corresponds to the \comsb\, shown in Fig.~\ref{Fig6}$a$.
Of importance here is that the total degassing flux at the \comsb\, is always higher than the supply limit for $M_{\rm oc} > 70M_{\rm oc, \oplus}$.
Thus, for $M_{\rm oc} > 70M_{\rm oc, \oplus}$, the planet has no equilibrium climate (i.e., extremely hot or cold climate) because of the enhanced seafloor weathering and the supply limit.
% \sakujo{
% Note that the solid part of the green line is the same as that shown in Fig.~\ref{Fig6}$a$.
% The dashed green line is drawn in the following way:
% For given $T_{\rm s}$, we find the crossover point between the adiabat and melting curve (see section~\ref{sec:OSM}) to determine the ocean mass and seafloor temperature. Using the value of $T_{\rm floor}$ in Eq.~(\ref{SW}), we calculate $F_{\rm SW}$. In the equilibrium states, $F_{\rm D} + F_{\rm M} = F_{\rm SW}$.
% In doing so, we use the minimum value of $f_{\rm oc}$ given in Fig.~\ref{Fig3}, which means the estimated value is a lower limit.
% }

% \sakujo{
% Thus, consequently, the total degassing flux has a peak at $128M_{\rm oc, \oplus}$.
% }

Here we give a brief discussion about the uncertainty in the supply limit.
Although the mean mantle heat flow $\bar{q}$, which determines the seafloor spreading rate $A_{\rm S}$ and thus the supply limit $F_{\rm SW}^{\rm limit}$, decreases with age during planetary evolution,
its decrement on a timescale of billion years is known to be similar to the mean mantle flow for Earth-like planets with age of several billion years \citep[]{McG1989}.
Also, we have adopted the solidus temperature of rock for $T_{\rm sol}$ in Eq.~(\ref{semiinf}), instead of the potential temperature of the mantle, which leads to overestimating the supply limit approximately by a factor of 2 in the case of $T_{\rm sol} = 2000$~K, which corresponds to the potential temperature at hot initial states \citep[e.g., ][]{Taj1992}.
In addition, in the equilibrium states, the effects of variation in seafloor spreading rate are canceled out, because both of the supply limit and degassing flux have a linear dependence on seafloor spreading rate (Eqs.~[\ref{degassing}] and [\ref{supply-limit}]).
Thus, the uncertainty in $\bar{q}$ has a small influence on the climate diagram for ocean planets.

%\naka{(海洋水量と関係するSupply limitの不定性について追加)}
The hydrothermal carbonation depth $d_{\rm hy}$ would depend on ocean mass.
Some experiments suggest that the hydrothermal carbonation depth decreases with increasing seafloor pressure because thermal cracking becomes weaker \citep[]{Van2007}.
Thus, the supply limit is expected to decrease with ocean mass, which would extend the domain of the extremely hot climate in Fig.~\ref{Fig7}.

In conclusion, the enhanced seafloor weathering due to the formation of the sorbet region and the supply limit narrow the range of ocean mass of terrestrial planets with the equilibrium climates.
This implies the difficulty of clement climates, like the present Earth, on ocean planets with plenty of water.

%===============================================
%% Uncertainty of the HP ice
\subsection{Caveats}
\subsubsection{Ocean Layer Model} \label{sec:uHP}
Here we discuss the validity of our assumptions regarding the ocean layer, which include:
(1) No boundary layer exists at the top of the HP ice layer;
(2) The carbon partitioning between the atmosphere and ocean is always in equilibrium and the CO$_2$ content is constant through the sorbet region;
and
(3) The heat transport occurs in the vertically one dimension.

\begin{itemize}
	\item[(1)]
	Regarding convective transport in the HP ice, we have considered the presence of a thermal boundary layer at the bottom, but not at the top.
	To evaluate the effect of the top boundary layer (TBL) on the effective weathering area $f_{\rm oc}$,
	we have calculated $f_{\rm oc}$ in the same settings as in \citet[]{Fu2009}, who considered TBL in addition to a bottom boundary layer (BBL).
	Then, we have found that TBL leads to reducing the effective weathering area in the low surface temperature domain for a given ocean mass (e.g., $T_{\rm s} \lesssim 390$~K for $M_{\rm oc} = 200M_{\rm oc, \oplus}$, see also  Fig.~\ref{Fig2}$f$).
	This is because BBL is cooler and thicker without TBL than in the case with TBL.
	As discussed in section~\ref{sec:DST}, in this domain, the reduced thickness of BBL, $\delta$, increases the critical heat flow (Eq.~[\ref{q_cr}]) and, thus, reduces the effective weathering area (Eq.~[\ref{criticalheat}]).
	However, we have also found that the presence of TBL brings about little change in the maximum of $q_{\rm cr}$ at a given ocean mass (Fig.~\ref{Fig3}).
	This is because the same temperature gradient in BBL is achieved by a change in $T_{\rm s}$, given that
	TBL is assumed to follow the melting line of ice \citep[see Fig.2 in][]{Fu2009}.
	Hence, the climate diagram for ocean planets is almost unaffected by the presence of TBL.

	\item[(2)]
	We have assumed that the CO$_2$ circulation in the sorbet region occurs efficiently enough that carbon partitioning between the atmosphere and ocean remains in equilibrium.
	However, the CO$_2$ circulation (not the seafloor weathering) limits the consumption of atmospheric CO$_2$,
	if being slower than the response of the carbon budget in the ocean-atmosphere system.
	The latter is controlled by regassing in the environments of interest in this study, although depending on degassing, in general \citep[]{Taj1992}; thus, its timescale is $\sim \tau/\beta$.

	The CO$_2$ circulation occurs in the following way:
	Aqueous CO$_2$ converts to CO$_2$ ice quickly within the HP ice \citep[]{Bol2013} and, then, the ${\rm CO_2}$ ice moves together with the HP ice.
	In the HP ice layer, since the upward sorbet flow transports mass (and heat), the HP ice sinks accordingly for mass conservation.
	Below we estimate the sinking speed of the HP ice and the overturn timescale of the HP ice layer from energy balance and mass conservation.

	When heat is transported by thermal diffusion and sorbet flow,
	the energy balance is expressed as
	\begin{equation}
		Q = - k \frac{dT}{dr} + \chi \rho_{\rm l} (L + C_{\rm P} \Delta T_{\rm HP})w_{\rm l}, \label{eq:ebalance}
	\end{equation}
	where $Q$ is the heat flux from the oceanic crust, $\chi$ is the melt fraction, $L$ is the latent heat of HP ice, and $\rho_{\rm l}$, $C_{\rm P}$, and $w_{\rm l}$ are the density, specific heat, and flow speed of liquid water in the sorbet, respectively.
	The first and second terms on the right side represent the thermal conduction and melt advection, respectively.
	Note that we have assumed that permeable flow of liquid dominates the sorbet flow and, namely, neglected upwelling solid flow, which results in underestimating the sinking flux of the HP ice.
	From mass conservation and Eq.~(\ref{eq:ebalance}), the sinking speed of the HP ice, $w_{\rm HP}$, against the upwelling sorbet flow is given by
	\begin{equation}
		w_{\rm HP} = \frac{\chi \rho_{\rm l}}{(1-\chi) \rho_{\rm HP}} w_{\rm l} = \frac{Q+ k\frac{dT}{dr}}{(L+C_{\rm P}\Delta T_{\rm HP}) (1-\chi)\rho_{\rm HP}},
	\end{equation}
	where $\rho_{\rm HP}$ is the density of the HP ice.
	The thermal conduction flux along the melting curve is $- k dT/dr \approx 10$~mW~m$^{2}$ (see Fig.~\ref{Fig2}$b$).
	The heat flux from the oceanic crust of 80~mW~m$^{-2}$ being added,
	$Q + kdT/dr \approx 70$~mW~m$^{-2}$.
	The material properties of liquid water and HP ice are $\rho_{\rm HP}$ = 1400~kg~m$^{-3}$,
	$L$ = $4.2\times 10^5$~J~kg$^{-1}$ \citep[at 300~K from][]{Dun2010},
	and $C_{\rm P}$ = $4.1 \times 10^3$~J~kg$^{-1}$~K$^{-1}$ \citep[at 300~K from][]{Wai2007}.

	For terrestrial sea ice, permeability decreases abruptly for melt fraction below $\chi$ = 5~\% \citep[]{Gol1998}.
	Although not known well for the HP ice, we assume that the HP ice behaves in a similar way and use $\chi$ = 5~\%.
	According to our calculation results,
	$\Delta T_{\rm HP}$ = 78~K and $D$ = 99~km for $M_{\rm oc}$ = 100 $M_{\rm oc, \oplus}$ and $T_{\rm s}$ = 300~K.
	Then, the overturn timescale of the HP ice ($\equiv D/w_{\rm HP}$) comes out to be 44~Myr.
	Even for the $M_{\rm oc} = 200M_{\rm oc, \oplus}$, $D/w_{\rm HP}$ = 203~Myr.
	On the other hand, $\tau/\beta \sim 250$~Myr for the present Earth's condition, using the value of $\tau$ for the present Earth ($\sim 100$~Myr) \citep[]{Tur2002}.
	Thus, the CO$_2$ circulation occurs faster than the response of the carbon budget in the atmosphere-ocean system.

	The above estimate may remain to be refined.
	For example, using the relation between the Nusselt number ($Nu$) and the Rayleigh number, $Nu \propto Ra^{1/3}$, \citep[]{Tur2002}, the semi-infinite half-space cooling model shows that the residence timescale $\tau$ is $\propto \eta^{2/3}_{\rm man}$, where $\eta_{\rm man}$ is a viscosity of the mantle material, and, thus, depends strongly on mantle temperature and water content in the mantle.
	Indeed, the residence timescale is thought to have varied by an order of magnitude during the thermal evolution of the Earth \citep[]{Taj1992}.
	Also, seafloor weathering would be limited, if the planet has a thick HP ice and vigorous convective mantle.
	However, it is emphasized here that given a weak dependence of the \comsb\,
	even a sluggish circulation of ${\rm CO_2}$ in the HP ice could yield no significant change in \comsb.

	\item[(3)]
	We have assumed a vertically one-dimensional structure of the ocean and thus considered only vertical heat transport.
	In reality, the thermal structure of the ocean is more complicated because of convective patterns and inhomogeneous phase change.
	First, since the distance between the mid-ocean ridge and trench ($\gtrsim$ 10000~km) is much larger than the thickness of the HP ice (100~km), with which the size of convective cells is comparable \citep[e.g.,][]{Tur2002},
	detailed convective patterns matter little for the overall heat transfer in the HP ice layer.
	In addition, hydrodynamical simulations show a heat-pipe structure of the HP ice layer for high heat fluxes from the oceanic crust, which means phase change rarely occurs vertically throughout the ocean \citep[]{Cho2017, Kal2018}.
\end{itemize}

%===========================================================
\subsubsection{Atmospheric model} \label{sec:uAM}

Our climate modeling has demonstrated that
the runaway cooling due to atmospheric CO$_2$ drawdown generally occurs on ocean planets with plate tectonics, provided $M_{\rm oc} > \Mcrsb$.
Although we assume the runaway cooling ends up with the snowball state with $T_{\rm s}$ $<$ 273~K, 
it is to be examined more carefully whether the global snowball state is achieved or not.
Here we discuss some uncertainties of the atmospheric model.

We have assumed a constant surface albedo of 0.32.
Planetary albedo depends on cloud radiative forcing that generally depends on $T_{\rm s}$ \citep[e.g.,][]{Wol2013, Wol2015}.
As far as partially ice-covered planets are concerned, simulations based on 3-D general circulation models (GCMs) for the Archean Earth \citep[][]{Wol2013} and ocean planets \citep[][]{Cha2017} demonstrate that planetary albedo increases rapidly with  decreasing  $T_{\rm s}$ due to the ice-albedo feedback, although the contribution of clouds to planetary albedo declines. 
This means that the assumed surface albedo of 0.32 is an underestimate for the planetary albedo for cold climates of interest in this study and thereby leads to overestimating $T_{\rm s}$.
This indicates that the runaway cooling results in the snowball state, even if the planet receives high stellar insolation comparable to the present Earth.

Also, we have assumed that the distribution of relative humidity in the atmosphere is the same as that in the present Earth's atmosphere.
\citet{Wor2013b} and \citet{Ram2014} found multiple equilibrium solutions for a CO$_2$-free, almost water-saturated atmosphere, including a hot solution with the surface temperature of $\sim~500$~K, even if stellar insolation is comparable to the present Earth's stellar insolation.
This suggests that the snowball state is not always achieved.  
However, GCM simulations show that atmospheric circulation leads to precipitation and thereby to removing water vapor from the atmosphere, namely, making unsaturated regions even if stellar insolation is close to the runaway greenhouse limit \citep[]{Wol2015}. 
This suggests that such a hot state would be unlikely to occur, although more work is needed to confirm so.

%Although atmospheric mass and another greenhouse gas are also important for planetary climate \citep[e.g.,][]{Pie_Gai2011}, it is beyond the scope of this study.

%===========================================================
\subsubsection{Carbon cycle model} \label{sec:ucarbon}

In this study we have ignored the situation where the atmosphere is so cold that the surface of the ocean is frozen and, instead, have stopped calculations once the surface temperature reaches 273~K.
Here we discuss the effect of surface ice on the carbon cycle and warming process in the snowball state. 
When the surface ice is convectively stable, which is appropriate for Earth-like high heat fluxes and moderately low surface temperatures \citep[]{Fu2009},  
molecular diffusion in the surface ice would control the exchange of CO$_2$ between the atmosphere and ocean. 
Performing molecular dynamics simulations, \citet{Ike2004} estimated that CO$_2$ molecular diffusion coefficient in  H$_2$O ice is $\sim 10^{-10}$~m$^2$~s$^{-1}$ at 270~K. 
For the thickness of surface ice of 1~km, for example, the diffusion timescale is on the order of Gyr. 
This means that even for a degassing flux higher than the critical value shown in Fig.\ref{Fig6}, CO$_2$ accumulation in the atmosphere proceeds too slowly for the climate to escape from the snowball state. 
This indicates that the snowball state we have found is maintained on a timescale of Gyr. 
We have to keep in mind, however, that it still remains a matter of debate how past Earth escaped from the snowball state.

On the other hand, the seafloor weathering is thought to be insensitive to the existence of surface ice, as follows.
The seafloor temperature is fixed at the melting temperature of the ice in the snowball solutions.
Since surface ice has a steep conductive temperature gradient and thus the thickness is small relative to the whole ocean, the $T$-$P$ structure below the surface ice is rather insensitive to the surface temperature, 
%\ikome{$\leftarrow$ DOESN'T MAKE SENSE. RATHER, I FEEL STEEP TEMPERATURE GRADIENT AFFECTS THE TEMPERATURE AT THE BOTTOM OF SURFACE ICE LAYER. BY THE WAY, ONCE THE OCEAN MASS IS GIVEN, ISN'T SEAFLOOR PRESSURE FIXED?}
thereby having little effect on the seafloor pressure and temperature. 
Thus, the seafloor weathering flux is proportional to the effective weathering area.
The latter increases with decreasing the surface temperature in the low surface temperature regime shown in Figure~\ref{Fig2}$f$. 
Thus, beyond the critical ocean mass, the seafloor weathering would be higher than the degassing flux, even if the surface ice is formed.
Thus, once being achieved by the runaway cooling, the snowball state is maintained.

%===============================================
%% Exoplanet
\subsection{Exoplanet} \label{sec:EXO}
Finally, we discuss an application of our findings to terrestrial exoplanets.
Although we have no enough knowledge of the degassing flux of exoplanets, which depends on several uncertain factors such as planetary carbon budget, thermal structure of planetary interior, and ocean mass,
we have found that the \comsb\, is less sensitive to the degassing flux.
As shown in Fig.~\ref{Fig6}$a$, terrestrial exoplanets with oceans of more than several tens of $M_{\rm oc, \oplus}$ in the habitable zone have extremely cold climates.
Cold climates are also suggested for Earth-like planets with low degassing flux in the habitable zone \citep[e.g.,][]{Kad2014}.
Thus, terrestrial planets with ${\rm CO_2}$-poor cold climates would not be uncommon in the habitable zone around Sun-like stars, provided plate tectonics is common for those planets.

Recently, habitability for planets around ultra cool stars (e.g., Proxima Centauri and TRAPPIST-1) are actively debated \citep[e.g.,][]{Rib2016, Tur2018, Valencia2018}.
Since the snowline is located near the habitable zone and ice-rich planets readily migrate from beyond, ocean planets would be abundant in the habitable zone around cool stars \citep[e.g., ][]{Tia2015}.
%\ikome{I DON'T RECOMMEND TO ADD NEW RESULTS IN DISCUSSION SECTION. IF YOU WANT TO SHOW THAT RESULT, YOU HAVE TO PROVIDE HOW TO CALCULATE IN DETAIL ENOUGH FOR THE READER TO REPRODUCE$\rightarrow$}
%\sakujo{
%Cool stars emit their radiation at a longer wavelength, which can strongly affect the planetary albedo \citep[e.g.,][]{Shi2013}. We performed the additional simulation for the planet around cool stars, using M4.5V star AD Leo spectrum \citep[]{Seg2005}. The simulation is performed for $S = 0.8 S_\oplus$ because the runaway greenhouse limit for cool stars is lower than the Sun's \citep[]{Kop2013}. We confirmed that the spectrum shift has no influence on the surface temperature because the spectrum shift does not alter the trend of $T_{\rm s}$ as a function of $P_{\rm CO_2}$. Thus, spectrum shift only affects $P_{\rm CO_2}$ in equilibrium and snowball states, as discussed in \S~\ref{sec:DSL} for dependence on the stellar insolation. Therefore, the spectrum shift has little influence on our results.
%}
%
Planets around cool stars are synchronously rotating, which results in a large difference in surface temperature between the day and night sides \citep[e.g.,][]{Pie2011}.
This might result in different phase structure and flow pattern in the ocean layer from our situation and ${\rm CO_2}$ condenses on the night side \citep[]{Tur2018}.
In this case, the HP ice would be easily formed on the cool night side.

However, provided all our assumptions are valid also for synchronously rotating planets and the dayside and nightside have the same thickness of the ocean layer, the weathering flux on the dayside is always higher than that on the nightside, because of high surface temperature due to the concentration of all the stellar insolation.
Thus, an equilibrium climate could be achieved on the dayside, although the nightside is extremely cold.
Then, the \comsb\, for such a planet can be defined in the same way as we have done above and its value is equivalent to the estimate given in the previous sections.
This implies a low probability of exoplanets with temperate climates in the habitable zone also around cool stars.
Note that even if they have massive oceans with a mass larger than the \comsb, synchronously rotating planets never become snowballs, because the local climate around the substellar point could be always temperate \citep[]{Che2017}.
\section{Summary and Conclusion}
The Earth's climate is stabilized by temperature-dependent, efficient continental weathering. Beyond the solar system, however, there must be continent-free terrestrial planets covered with global oceans (called ocean planets).
Only with inefficient seafloor weathering, the Earth's climate would be much warmer.
Furthermore, previous studies suggest that ocean planets have extremely hot climates, if they have massive oceans of 20 to $\sim 100 M_{\rm oc, \oplus}$,  because the HP ice present in the deep ocean completely prevents chemical weathering on the oceanic crust \citep[][]{Ali2014,Kit2015}.
However, those studies oversimplify the heat transfer in the HP-ice layer and ignore horizontal variation from heat flow from the oceanic crust.
Thus, in this study, we have revisited the climate of ocean planets with plate tectonics in the habitable zone, by incorporating the effects of the liquid/solid coexistence region (called the sorbet region) near the mid-ocean ridge in the carbon cycle (Fig.~\ref{fig1}).
The main findings of this study are summarized as follows.

Our seafloor environment model without the effect of the carbon cycle (i.e., fixed surface temperature) has shown that even if pressures in the deep ocean are high enough for HP ice to form, heat flux from the crust is too high to be transferred by solid convection, making the HP ice molten and forming a sorbet region, at least, near the mid-ocean ridge (section~\ref{sec:melting}).
Although reduced with increasing ocean mass or decreasing mean mantle heat flow, the effective weathering area never becomes zero for $M_{\rm oc} \leq 200 M_{\rm oc, \oplus}$.
This means that seafloor weathering remains possible and subsequent material circulation (e.g., carbon cycle) will sufficiently occur through the sorbet region.

Modeling the carbon cycle with the effect of seafloor weathering under the sorbet region,
we have found that the climate on the ocean planet is destabilized and lapses into a CO$_2$ poor, extremely cold state, which is called the snowball state (section~\ref{sec:ESW}).
Such destabilization is triggered because seafloor temperature is fixed at the melting temperature of the HP ice and, thus, a high seafloor weathering flux is kept regardless of surface temperature, unlike continental weathering which is dependent on surface temperature.
This indicates the existence of a critical ocean mass, beyond which an ocean planet no longer maintains a temperate climate.
We have demonstrated that the critical ocean mass is less sensitive to planetary mass, degassing flux, and the detailed dependence of seafloor weather flux on seafloor temperature (i.e., the activation energy $E_\mathrm{a}$), and is several tens of $M_{\rm oc, \oplus}$.
Also, because of the supply limit of cations, seafloor weathering is ineffective in compensating massive degassing, not achieving equilibrium climates, but yielding extremely hot ones.

As demonstrated in this paper, thermal and chemical interaction between the ocean and rocky interior significantly alters the planetary climate of ocean planets even in the habitable zone.
We have found that temperate equilibrium climates are achieved in limited ranges of ocean mass and degassing flux. 
This suggests that a certain proportion of terrestrial exoplanets in the habitable zone could be frozen ocean planets, provided they are Earth-like ones with plate tectonics. 
In any case, our findings indicate that ocean mass has a crucial role in the planetary climate of terrestrial planets with a massive ocean.
While the characterization of terrestrial exoplanets will be performed for detecting habitable planets in the next decade,
we should discuss their climates carefully because those exoplanets would be diverse in surface water amount.

%*************************************** Conclusion **********************************************

% %*************************************** Acknowledgements **************************
\section*{Acknowledgements}
The authors thank the anonymous reviewer for thoughtful comments that greatly improved the manuscript.
This work was supported by JSPS KAKENHI No.~JP18H05439, JP23103003, and JP17H06104, ABC-NINS No.~AB301002, JSPS Core-to-core Program ``{}International Network of Planetary Sciences''{}, and the European Research Council (ERC) under the European Union's Horizon 2020 research and innovation programme (grant agreement No.~679030/WHIPLASH).

% The Acknowledgements section is not numbered. Here you can thank helpful
% colleagues, acknowledge funding agencies, telescopes and facilities used etc.
% Try to keep it short.

%*************************************** References **********************************************
\bibliographystyle{mnras}
\bibliography{reference} % if your bibtex file is called example.bib

% %*************************************** Appendix **********************************************

%*************************************** Appendix **********************************************
\appendix

%===============================================
%% Internal structure model
\section{Internal structure model} \label{sec:ISM}

We develop a radially one-dimensional, hydrostatic internal structure model in section~\ref{sec:OSM}, based on \citet[]{Val2007}.
We consider a differentiated solid rock-metal body of 1 Earth mass covered with various amounts of ${\rm H_2O}$.
Note that the planetary mass is the sum of the rock-metal body and ocean masses (i.e., $1 M_\oplus + M_{\rm oc}$).
We assume (1) the mass ratio of iron core to silicate-mantle is 7~:~3, (2) the mass ratio of the inner to outer core is the same as that of the Earth (35~:~65), (3) phase transitions occur at the same pressures as in the Earth's interior, and (4) thermal expansion of the mantle and core never occurs.
These assumptions have little influence on the surface gravity, which affects the structure of the ocean layer and thus on our conclusion in this study.

The equations of state (EOSs) and parameter values that we adopt are summarized in Table~\ref{Tablea1}.
The temperature effect on density of the HP ice follows expressions from \citet[]{Bez2014} for ice~\rom{6} and \citet[]{Fei1993} for ice~\rom{7}.
The thermal capacity of liquid water is taken from \citet[]{Wai2007}.
The phase transition from water to HP ice occurs where the adiabat crosses the phase boundaries in the $P$-$T$ plane given by  Eq.~\ref{eq:Tmel}.

We assume the thermal structure of the iron core and the rocky mantle are the same as the Earth.
The phase transitions and the pressures at the transitions are summarized in Table~\ref{Tablea2}.

\begin{table*}
	\begin{center}
		\caption{Data for EOS parameters \label{Tablea1}}
		\begin{threeparttable}
		\begin{tabular}{c|c|ccccc} \hline
			Layer & Composition & $\rho_0$ & $B_0$ & $B'_0$ & EOS & Reference \\
			& & (kg ${\rm m^{-3}}$) & (GPa) & & & \\ \hline
			${\rm H_2O}$ & Liquid water & & & & & (1) \\
			& Ice~\rom{6} & 1270 & 14.05 & & BME & (2) \\
			& Ice~\rom{7} & 1240 & 5.02 & 7.51 & Vinet & (3) \\ \hline
			Upper mantle & ol & 3347 & 126.8 & 4.274 & Vinet & (4) \\
			& wd + rw & 3644 & 174.5 & 4.274 & Vinet & (4) \\ \hline
			Lower mantle & pv + fmv & 4152 & 223.6 & 4.274 & Vinet & (4) \\
			& ppv + fmv & 4270 & 233.6 & 4.524 & Vinet & (4) \\ \hline
			Outer core & ${\rm Fe_{0.8}(FeS)_{0.2}}$ & 7171 & 150.2 & 5.675 & Vinet & (4) \\ \hline
			Inner core & Fe & 8300 & 150.2 & 5.675 & Vinet & (4) \\ \hline
		\end{tabular}
		\begin{tablenotes}
			\item{
			$\rho_0$ is the reference density, $B_0$ is the bulk modulus and $B'_0$ is the pressure derivative of the bulk modulus.
			BME represents second-order Birch-Murnaghan EOS \citep[ref.][]{Bir1978}, Vinet is represented Vinet EOS \citep[ref.][]{Vin1989}.}
			\item{
			(1) \citet[]{Lev2014} and references therein; (2) \citet[]{Bez2014}; (3) \citet[]{Sug2008}; (4) \citet[]{Val2007} and references therein.}
		\end{tablenotes}
	\end{threeparttable}
	\end{center}
\end{table*}

\begin{table}
	\begin{center}
		\caption{Phase boundaries of rocky material \label{Tablea2}}
		\begin{threeparttable}
		\begin{tabular}{ccc} \hline
			Phase transition &  Boundary pressure & Reference\\ \hline
			ol $\rightarrow$ wd + rw & 13.5 GPa & (1) \\
			rw $\rightarrow$ pv + fmv & 23.1 GPa & (1) \\
			pv + fmv $\rightarrow$ ppv + fmv & 125 GPa & (2) \\
			\hline
		\end{tabular}
		\begin{tablenotes}
			\item{(1)\citet[]{Tur2002}; (2)\citet[]{Mur2004}}
		\end{tablenotes}
	\end{threeparttable}
	\end{center}
\end{table}

%===============================================
\section{Partitioning of ${\rm CO_2}$ between atmosphere and ocean} \label{sec:patitioning}
The partial pressure of ${\rm CO_2}$, $P_{\rm CO_2}$, depends on the carbon budget of the surface reservoirs ($C_{\rm atm} + C_{\rm oc}$), oceanic pH and ocean volume, $V_\mathrm{oc}$.
Here we describe the calculation method of carbon partitioning between the atmosphere and ocean, which is almost the same as that described in \citet[]{Taj1990} and \citet[]{Kit2015}.

Chemical equilibrium among ${\rm CO_2 (g)}$, carbonic acid ($\rm H_2CO_3$), carbonate ion ($\rm CO_3^{2-}$), and bicarbonate ion ($\rm HCO_3^{-}$) is determined by the following reactions
\begin{eqnarray}
	{\rm CO_2 (g) + H_2O} &\rightleftharpoons& {\rm H_2CO_3} , \label{eq:henry}\\
	{\rm H_2CO_3} &\rightleftharpoons& {\rm HCO^{-}_3 + H^+} , \\
	{\rm HCO_3^-} &\rightleftharpoons& {\rm CO^{2-}_3 + H^+} , \\
	{\rm H_2O} &\rightleftharpoons& {\rm H^+ + OH^-} , \label{eq:water}
\end{eqnarray}
with equilibrium constants
\begin{eqnarray}
	K_0 &=& \frac{\rm [H_2CO_3]}{P_{\rm CO_2}} , \label{eq:K0}\\
	K_1 &=& \frac{\rm [H^+][HCO^-_3]}{\rm [H_2CO_3]} , \\
	K_2 &=& \frac{\rm [H^+][CO^{2-}_3]}{\rm [HCO^-_3]} , \\
	K_{\rm w} &=& {\rm [H^+][OH^-]}. \label{eq:Kw}
\end{eqnarray}
We have assumed the Henry's law is valid. We use the values of $K_0$ from \citet[]{Wei1974}, $K_1$ and $K_2$ from \citet[]{Mil2006}, and $K_{\rm w}$ from \citet[]{Mil1995}, which were obtained experimentally for temperature of 300~K and salinity of 35~\textperthousand. We use those values at all temperatures, because those equilibrium constants have not been experimentally measured for the temperature range of interest in the study.

Since Equations~(\ref{eq:K0})--(\ref{eq:Kw}) contain six unknowns, we need at least two additional equations.
One is the equation of charge conservation:
\begin{equation}
 	{\rm [OH^-] + [HCO^-_3] + 2[CO^{2-}_3] = [H^+] + [M^+],}
 \end{equation}
where [M$^+$] represents the concentration of all the cations in the ocean.
In this study, we use the average value of $\rm [M^+]$ measured in the present Earth ocean (i.e., $\rm [M^+]$ $= 2.2 \times 10^{-3}$~mol~L$^{-1}$).
Even a constant value of $\rm [M^+]$ has little influence on conclusions (see also section~\ref{sec:carbon}).
Also, the total number of carbon $C_\mathrm{atm}+C_\mathrm{oc}$, which is determined from the carbon cycle model (section~\ref{sec:carbon}), must be conserved in the atmosphere-ocean system.
Since the solubility of ${\rm CO_2}$ in liquid water increases rapidly with pressure \citep[]{Dua2003} and mixing occurs in the ocean on a timescale much shorter than that of interest in this study, we assume $C_\mathrm{oc}$ to be equal to the surface concentration of ${\rm CO_2}$.
Thus, the total number of carbon is expressed as
\begin{equation}
	C_\mathrm{atm} + C_\mathrm{oc} \approx
	\frac{4 \pi R_{\rm p}^2 P_{\rm CO_2}}{m_{\rm CO_2} g_{\rm s}} +
	( {\rm [H_2CO_3] + [HCO^-_3] + [CO^{2-}_3]} ) V_{\rm oc}, \label{eq:R_oc}
\end{equation}
where the first term on the right-hand side corresponds to $C_\mathrm{atm}$, and $m_\mathrm{CO_2}$ is the molecular weight of ${\rm CO_2}$ (= $44~{\rm g}~{\rm mol}^{-1}$) and $g_{\rm s}$ is the surface gravity.
We obtain $R_{\rm p}, g_{\rm s}$, and $V_{\rm oc}$ for given $T_{\rm s}$ and $M_{\rm oc}$ from the internal structure model in section~\ref{sec:OSM}.
For $C_\mathrm{atm}$, we assume that the molecular weight of the atmospheric gas is equal to the molecular weight of $\rm CO_2$, which overestimates $P_{\rm CO_2}$.
However, the approximation has no influence on the overall results of the study.

Finally, solving Eqs.~(\ref{eq:K0})-(\ref{eq:R_oc}), we determine $P_\mathrm{CO_2}$ and the mole fractions of ions.

%*************************************** Appendix **********************************************
\section{Dependence of degassing coefficient on seafloor pressure} \label{sec:DRM}

Here we introduce the dependence of degassing coefficient, $K_{\rm D}$, on seafloor pressure.
According to \citet[]{Taj1992}, the degassing coefficient is given by
\begin{equation}
	K_{\rm D} = f_{\rm CO_2} \frac{d_{\rm m}}{V_{\rm man}},
\end{equation}
where $f_{\rm CO_2}$ is the degassing fraction, which is defined as the molar fraction of the ${\rm CO_2}$ degased from the upwelling magma at the ridge, $d_{\rm m}$ is the degassing depth, which is defined as the melt generation depth of mantle, and $V_{\rm man}$ is the volume of the mantle.
For $V_\mathrm{man}$ and $d_\mathrm{m}$, we use the values for the present Earth, namely $V_\mathrm{man}$ $= 8.0 \times 10^{20}$~m$^3$ and $d_{\rm m} = 40$ km \citep[]{Taj1992}.

The degassing fraction $f_{\rm CO_2}$ depends on the ocean mass, because of pressure dependence of ${\rm CO_2}$ solubility into magma \citep[]{Kit2009}.
In this study, we take it into account, following \citet[]{Taj1992}, who considered the solubility equilibrium of ${\rm CO_2}$ with solid/liquid silicate.
We incorporate the pressure dependence on the solubility of ${\rm CO_2}$ into silicate melts, $K^{\rm G/L}$, and the molar volume of CO$_2$, $V_{\rm M}$, in calculating $f_{\rm CO_2}$ as \citep[]{Taj1992}
\begin{equation}
	f_{\rm CO_2} = \left\{ 1+\frac{(f_{\rm melt}^{-1}-1)K^{\rm L/S}}{1+(w^{\rm G}/w^{\rm L})} \right\}^{-1} , \label{eq:fco2}
\end{equation}
where $f_{\rm melt}$ is the melt fraction, $K^{\rm L/S}$ is the partition coefficient of ${\rm CO_2}$ between solid and liquid, and $w^{\rm G}/w^{\rm L}$ represents the mass ratio of ${\rm CO_2}$ partitioned into the gas phase to that into the liquid phase (liquid phase meaning ${\rm CO_2}$ dissolved in melt);
$w^{\rm G}/w^{\rm L}$ is defined as
\begin{equation}
	 \frac{w^{\rm G}}{w^{\rm L}} = \frac{\phi m_{\rm CO_2} n_{\rm CO_2}}{\rho_{\rm rock} V_{\rm M} K^{\rm G/L}} ,
\end{equation}
where $\phi$ is the vesicularity of melt, $\rho_{\rm rock}$ is the density of oceanic crust, and $n_{\rm CO_2}$ is the molar concentration of ${\rm CO_2}$ gas in the vesicles.
We adopt values of $f_{\rm melt}, K^{\rm L/S}, \phi$, and $n_{\rm CO_2}$ from \citet[]{Taj1992}

Recent molecular dynamics simulations \citep[]{Gui2011} predict higher solubility of ${\rm CO_2}$ for $> 2$~GPa than that obtained according to Henry's law.
Those simulations found an almost linear dependence on pressure and weakly correlation with temperature.
We have estimated the relationship between $K^{\rm G/L}$ and $P$ based on tabular data for $T=$ 1673~K and MORB composition presented in \citet[]{Gui2011}:
\begin{eqnarray}
	K^{\rm G/L} = \left\{
	\begin{array}{ll}
		0.008 P  & P < 2{\rm GPa}\\
		0.035 (P - 2.0) + 0.016 ~~~ & P \geq 2 {\rm GPa}  .
	\end{array}
	\right.
\end{eqnarray}
Here $P$ is the pressure in the unit of GPa.
We evaluate $V_{\rm M}$ at the seafloor pressure using the EOS based on molecular dynamics simulations \citep[]{Dua2006}, which is of wide application (i.e., up to 10~GPa and 2573.15~K).
The temperature in the EOS corresponds to the solidus of anhydrous peridotite at the seafloor pressure, which is parameterized by \citet[]{Hir2009}.

\begin{figure}
	\includegraphics[width=\columnwidth]{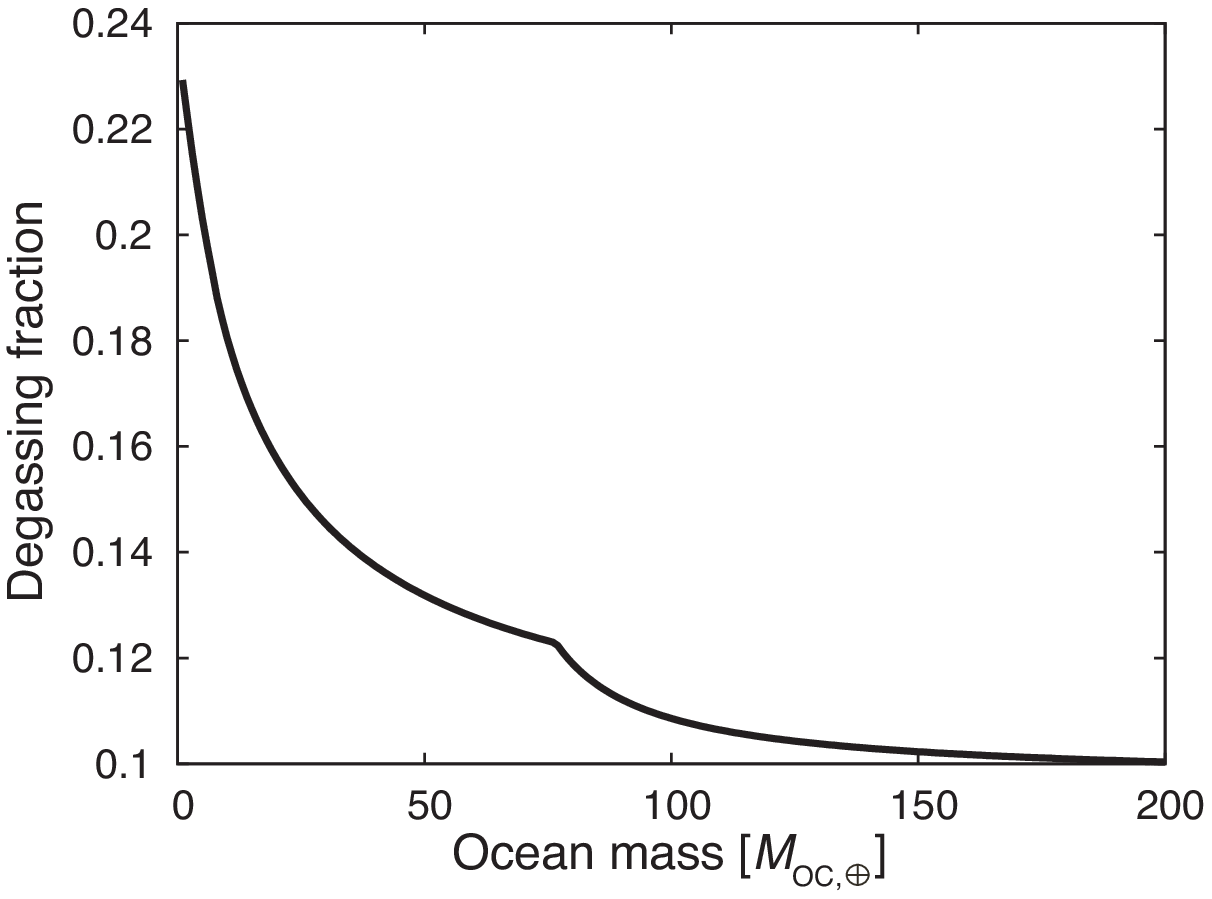}
	\caption{
	Degassing fraction $f_{\rm CO_2}$ (see Eq.[\ref{eq:fco2}]) as a function of ocean mass for surface temperature $T_{\rm s} = 300$~K. $M_{\rm oc, \oplus}$ means the Earth's ocean mass.
	\label{Figa2}}
\end{figure}

Figure~\ref{Figa2} shows the degassing fraction $f_{\rm CO_2}$ as a function of ocean mass $M_\mathrm{oc}$ for $T_{\rm s} = 300$~K.
$f_{\rm CO_2}$ is found to decrease with $M_\mathrm{oc}$, because the solubility of CO$_2$ increases with pressure.
$f_{\rm CO_2}$ varies from 0.23 to 0.1 between 1 to 200~$M_{\rm oc, \oplus}$.
At $M_\mathrm{oc}$ = 77~$M_{\rm oc, \oplus}$, the slope of $f_{\rm CO_2}$ changes because seafloor pressure becomes higher than 2~GPa.
\citet[]{Kit2009} proposed that degassing could be completely suppressed (i.e., $f_{\rm CO_2} = 0$) for a 100~km ocean (roughly 40~$M_{\rm oc, \oplus}$ in our model) because of higher solubility of ${\rm CO_2}$.
In contrast, Fig.~\ref{Figa2} indicates that degassing also occurs for larger $M_\mathrm{oc}$.
Higher solubility would lead to no partitioning into the gas phase ($w^{\rm G}/w^{\rm L} \rightarrow 0$).
In this case, degassing fraction would be determined by two-phase partitioning between solid and liquid and consequently $f_{\rm CO_2}$ becomes 0.096.
Therefore, our model results in degassing that mainly occurs as the liquid phase at high pressures.

Also, in Fig.~\ref{Figa2}, $f_{\rm CO_2}$ is estimated to be 0.23 for $M_\mathrm{oc}$ = 1~$M_{\rm oc, \oplus}$ corresponding to seafloor pressure of 27~MPa, which is relatively smaller than the value (0.32) estimated according to the Henry's law, by \citet[]{Taj1992}.
This difference is due to higher solubility (216~ppm at 27~MPa) than that (100~ppm) of \citet[]{Taj1992}. Note that low-pressure experiments suggest higher solubility than our model \citep[]{Jen1997}.
In any case, because a pressure range much higher than 27~MPa is of special interest in this study, we neglect this difference.

\bsp	% typesetting comment
\label{lastpage}

\end{document}